\documentclass[aps,prb,amsmath,amssymb,footinbib,showpacs,twocolumn,superscriptaddress]{revtex4-2}
\usepackage{amsmath}
\usepackage{amssymb}
\usepackage{amsthm}
\usepackage{setspace}
\usepackage{graphicx}
\usepackage{braket}
\usepackage{mathrsfs}
\usepackage{times}
\usepackage{bm}
\usepackage[linktocpage=true,colorlinks=true,urlcolor=blue,citecolor=red]{hyperref}
\usepackage{pgfplots}
\usepackage{amsfonts}
\usepackage{amsmath}
\usepackage{amssymb}
\usepackage{amsthm}
\usepackage{mathrsfs}
\usepackage{setspace}
\usepackage{xcolor}
\usepackage{braket}
\usepackage{relsize}
\usetikzlibrary{arrows}
\usepackage{enumerate}
\usepackage{float}

\begin{document}

\title{Fermion parity and quantum capacitance oscillation with partially separated Majorana and quasi-Majorana modes}

\author{Tudor D. Stanescu}
\affiliation{Department of Physics and Astronomy, West Virginia University, Morgantown, WV 26506, USA}
\author{Sumanta Tewari}
\affiliation{Department of Physics and Astronomy,
Clemson University, Clemson, SC 29634}

\begin{abstract}
In a recent experiment, flux-dependent oscillations of the quantum capacitance were observed in a one-dimensional spin-orbit-coupled semiconductor–superconductor heterostructure connected end-to-end via a quantum dot and threaded by a magnetic flux. In the topological superconducting phase of the heterostructure, the oscillations corresponding to different fermion parity sectors are shifted by half a period and can serve as a mechanism for fermion parity readout or fusion operations involving a pair of localized, well-separated Majorana modes. In this work, we demonstrate that flux-induced fermion parity-dependent oscillations of the quantum capacitance in a disordered semiconductor–superconductor-quantum dot system can originate not only from topologically protected, spatially well-separated Majorana zero modes (MZMs) localized at the wire ends, but also, generically, from partially-separated Majorana modes with significant overlap, as well as from quasi-Majorana modes in the topologically trivial phase, which can be viewed as Andreev bound states whose constituent Majorana wave functions are slightly shifted relative to each other and have non-zero amplitude at opposite ends of the wire. Therefore, while the detection of flux-dependent oscillations of quantum capacitance marks an important experimental advance, such observations alone do not constitute evidence of the presence of topological Majorana zero modes. 
\end{abstract}

\maketitle

\section{Introduction} 
Majorana zero modes (MZMs) \cite{Read2000,Kitaev_2001, Kitaev2003}, the condensed matter counterparts of Majorana fermions from high-energy physics \cite{Majorana1937}, can be understood as resulting from a non-local fractionalization of a conventional fermion mode into a pair of Majorana zero modes \cite{Read2000,Kitaev_2001}. Due to their capacity to encode quantum information non-locally and their associated exotic statistics -- known as non-Abelian statistics \cite{wilczek1982quantum, Moore1991, Read2000, Nayak1996} -- they have been proposed as fundamental building blocks for topological quantum computation (TQC) \cite{Kitaev_2001, Kitaev2003, Nayak2008}. It has been proposed that MZMs can be realized experimentally in a one-dimensional semiconductor-superconductor (SM-SC) heterostructure subjected to a parallel Zeeman field
\cite{sau2010generic, sau2010non, oreg2010helical, lutchyn2010majorana}. When the field exceeds a critical value, the system is predicted to transition into a one-dimensional topological superconducting (1DTS) phase that hosts MZMs localized at the ends, leading to extensive experimental efforts and several promising results in recent years \cite{mourik2012signatures, Deng2012, Das2012, rokhinson2012fractional, churchill2013superconductor, finck2013anomalous, deng2016majorana, zhang2017ballistic, chen2017experimental, nichele2017scaling, albrecht2017transport, o2018hybridization, shen2018parity, 
sherman2017normal, vaitiekenas2018selective, albrecht2016exponential, Yu_2021, zhang2021, PhysRevB.107.245423}. 
However, it has also become evident that disorder can induce trivial low-energy states that often imitate the signatures of topological MZMs \cite{kells2012near, bagrets2012class, pikulin2012zero, prada2012transport, Liu_2017, pan2020physical, moore2018two, Moore2018, vuik2018reproducing, stanescu2019robust, added_Loss_2018prb_abs, san2016majorana, ramon2019nonhermitian, Jorge2019supercurrent, ramon2020from, Jorge2021distinguishing}. Notably, the so-called partially-separated Andreev bound states (ps-ABS) or quasi-Majorana zero modes (q-MZMs) \cite{moore2018two,Moore2018,vuik2018reproducing} are ABS states whose constituent Majorana wave functions are spatially separated just enough to form robust low-energy states resembling MZMs, but not sufficiently to reduce their overlap to a level that would make the states insensitive to local perturbations, i.e., a separation that would provide topological protection [see, e.g., Fig. \ref{FIG6}, panel (a)].
Such states are generically induced by strong-enough disorder in the topologically trivial phase of the heterostructure [see, e.g., Fig. \ref{FIG5}, panel (a)].  

The clean or weakly-disordered 1D SM-SC heterostructure with applied Zeeman field above a certain critical value is a system in which the lowest-energy excitations consist of a pair of MZMs localized at the wire ends. MZMs differ from conventional fermionic modes in that they do not possess a well-defined fermion occupation number. To define a conventional fermionic state in a system hosting MZMs, one 
must consider a pair of such modes. For example, the pair of MZMs $\gamma_1$ and $\gamma_2$, localized at the ends of a SM–SC hybrid nanowire in the topological phase [see, e.g., the schematic illustration in Fig. \ref{FIG1}(a)] can be combined into 
a zero-energy complex fermion mode represented by the operator $c^{\dagger} = \frac{1}{2}(\gamma_1 + 
i\gamma_2)$. The charge state of the system is then given by the eigenvalue of $n_c = c^{\dagger}c = 0, 1$. 
The fermion parity $F = (-1)^{n_c}$ associated with the operator $c^{\dagger}$ is related to 
the MZM operators $\gamma_1, \gamma_2$ via $F = (-1)^{n_c} = i\gamma_1\gamma_2$, indicating that the 
system’s fermion parity is determined by non-local correlations between the spatially separated, 
fractionalized MZMs $\gamma_1$ and $\gamma_2$. In fact, this fractionalization of fermion parity into a pair of spatially separated MZM operators is a defining feature of the system’s topological nature \cite{Turner_2011}. An unambiguous measurement of the fermion parity shared by a pair of MZMs constitutes a fusion operation, which underlies measurement-only topological quantum computation (TQC) involving Majorana qubits \cite{Bonderson_2008}. However, the non-local nature of the fermion parity, arising from its fractionalization between two spatially separated MZMs, renders its 
definitive measurement inherently challenging \cite{Sau_Tewari_2015, Akhmerov_2009, Sau_Tewari_2011}. In a recent experiment \cite{Microsoft2025}, measurements of flux-dependent oscillations of the quantum capacitance in a one-dimensional spin-orbit-coupled SM-SC hybrid nanowire connected end-to-end via a quantum dot and threaded by a magnetic flux, were used for fermion parity readout or fusion operations involving a pair of localized MZMs in a putative Majorana qubit.

In this paper, we study a one-dimensional SM-SC hybrid nanowire with Rashba spin-orbit coupling (SOC) and a parallel magnetic field in the presence of intermediate-to-strong disorder. Reflecting the experimental setup, we assume that the two ends of the nanowire -- referred to as a Majorana nanowire in Fig. \ref{FIG1} -- are connected end-to-end via an extended quantum dot (QD), forming a loop which is threaded by a perpendicular magnetic flux $\Phi$. In the topological superconducting phase of the SM–SC nanowire, the coupling of the end MZMs, $M_1$ and $M_2$, through the quantum dot gives rise to a pair of finite-energy Andreev bound states (ABSs), described by $c^{\dagger}|0\rangle = \frac{1}{2}(\gamma_1 + i\gamma_2)|0\rangle$ and $c|0\rangle = \frac{1}{2}(\gamma_1 - i\gamma_2)|0\rangle$, where $|0\rangle$ denotes the superconducting ground state. These states possess opposite fermion parity, $\pm 1$, (i.e., $i\gamma_1\gamma_2 c^{\dagger}|0\rangle = -c^{\dagger}|0\rangle$, $i\gamma_1\gamma_2 c|0\rangle = +c|0\rangle$), and differ in the occupation number $n_c = 1, 0$ of the conventional fermionic mode formed by $\gamma_1$ and $\gamma_2$. The different occupation numbers of the ABS state in the quantum dot result in distinct values of the quantum capacitance of the SM–SC–QD system. Therefore, measuring the quantum capacitance provides indirect access to the fermion parity associated with the end MZMs. The quantum capacitance of the QD may depend on factors other than the fermion parity (i.e., the ABS occupation), such as the flux $\Phi$ generated by a perpendicular magnetic field threading the interferometric loop, but this dependence is parity-sensitive, providing a distinct signature of changes in the fermion parity \cite{Microsoft2025}.

We first calculate the flux dependence of the energies of the ABS pair ($c^{\dagger}|0\rangle$, $c|0\rangle$), corresponding to odd and even fermion parity, for a system with weak disorder. Analogous to the fractional Josephson effect \cite{Kitaev_2001}, we show that in the 1DTS phase of a weakly disordered $3 \mu$m SM-SC quantum wire, the energy–flux relation $E(\Phi)$ exhibits periodicity with period $h/e$, which is twice the conventional superconducting flux quantum $h/2e$. Correspondingly, the quantum capacitance -- computed using a linear response formalism \cite{Sau2024} -- is also periodic in $\Phi$ with the same $h/e$ period. These flux-periodic oscillations of the quantum capacitance in the 1DTS phase with localized MZMs are consistent with recent experimental observations in QD-SM-SC heterostructures \cite{Microsoft2025}.
We then extend our analysis to the case of intermediate-to-strong disorder, again for a $3 \mu$m quantum wire. Remarkably, we find that flux-dependent oscillations in fermion parity -- or, equivalently, in quantum capacitance -- can emerge not only from topologically protected MZMs localized at the ends of the wire [Fig. \ref{FIG2}, panel (a)], but also, quite generically, from partially separated MZMs that exhibit significant spatial overlap while still maintaining non-zero amplitude at opposite ends of the wire [e.g., Fig. \ref{FIG6}, panels (b), (c), (d)]. Even more significantly, we find that qualitatively similar parity-shifted, flux-dependent oscillations in capacitance can also arise from quasi-Majorana modes (q-MZMs), which are generically induced by disorder in the topologically trivial phase [Fig. \ref{FIG6}, panel (a)]. Consequently, while the observation of parity-dependent, flux-induced oscillations of the quantum capacitance in QD-SM-SC systems represents an important experimental advance, such oscillations alone do not constitute indisputable evidence of the presence of topological MZMs.

\section{Hamiltonian for the hybrid nanowire coupled to a quantum dot}  
\begin{figure}[t]
\begin{center}
\includegraphics[width=0.48\textwidth]{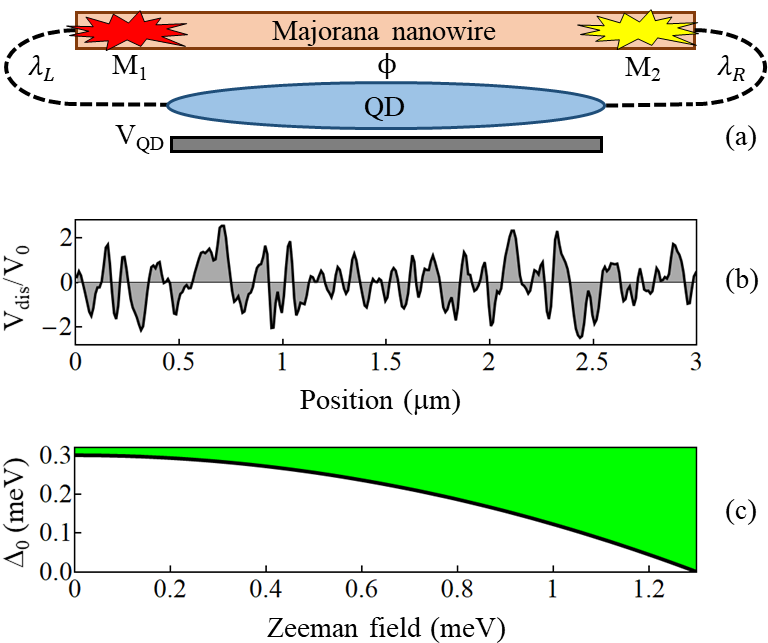}
\end{center}
\vspace{-2mm}
\caption{(a) Schematic representation of a semiconductor-superconductor hybrid nanowire coupled to a quantum dot (QD). The effective coupling at the left (right) end of the nanowire is $t\cdot\lambda_L$ ($t\cdot\lambda_R$), where $t$ is the nearest-neighbor hopping of the 1D tight-binding model describing the SM wire [see main text, Eqs. (\ref{HSM}) and (\ref{HQD})]. A back gate controls the potential $V_{QD}$ of the quantum dot, while a magnetic flux $\Phi$ threads the space between the wire and the dot. In the topological regime, the nanowire supports a pair of Majorana modes (M$_1$ and M$_2$) localized near the ends. (b) Position dependence of the (normalized) disorder potential in the nanowire. (c) Dependence of the parent superconducting gap on the Zeeman field applied parallel to the wire.}
\label{FIG1}
\vspace{0mm}
\end{figure}

We consider an interferometric loop consisting of a $3$ micron Majorana nanowire, e.g., a quasi-1D SM-SC hybrid structure, coupled to a quantum dot (e.g., a SM wire) \cite{Microsoft2025}. The device is represented schematically in Fig. \ref{FIG1}(a). A magnetic field $B$ applied parallel to the wire generates a Zeeman splitting $\Gamma=\frac{1}{2}g \mu_B B$, where $g$ is the Land\'{e} g-factor of the semiconductor and $\mu_B$ is the Bohr magneton, while a magnetic flux $\Phi$ threads the interferometric loop. The SM wire is modeled using a simple 1D tight-binding Hamiltonian and the proximity effect induced by the SC is captured, after integrating out the SC degrees of freedom, by a self-energy contribution. Specifically, the Green's function of the semiconductor has the form
\begin{equation}
G_{SM}(\omega) = [\omega - H_{SM} -\Sigma_{SC}(\omega)]^{-1}, \label{GSM}
\end{equation}
with $H_{SM}$ being the Bogoliubov-de Gennes matrix that corresponds to the second quantized Hamiltonian
\begin{eqnarray}
\hat{H}_{SM} &=& \sum_{i,\sigma}\left[-t(c_{i\sigma}^\dagger c_{i+1 \sigma}^{} + {\rm h.c.}) + (V_{dis}(i)-\mu)c_{i\sigma}^\dagger c_{i\sigma}^{}\right] \nonumber \\
&+& \frac{\alpha}{2}\sum_i\left[(c_{i\uparrow}^\dagger c_{i+1 \downarrow}^{}-c_{i\downarrow}^\dagger c_{i+1 \uparrow}^{}) + {\rm h.c.}\right] \label{HSM} \\
&+& \Gamma\sum_{i} (c_{i\uparrow}^\dagger c_{i+1 \downarrow}^{}+c_{i\downarrow}^\dagger c_{i+1 \uparrow}^{}), \nonumber
\end{eqnarray}
where $i=1,2,\dots,N$ (with $N=300$) labels the sites of a 1D lattice with lattice constant $a=10~$nm and $c_{i\sigma}^\dagger$ ($c_{i\sigma}$) designates the creation (annihilation) operator for an electron with spin $\sigma$ occupying site $i$. The effective parameters are the nearest-neighbor hopping $t=16.56~$meV, which corresponds to an effective mass $m^*=0.023m_0$ characterizing an InAs semiconductor wire, the chemical potential $\mu$, the Rashba spin-orbit coupling coefficient $\alpha=1.4~$meV (i.e., $\alpha\cdot a = 140~$meV$\cdot$\AA), and the Zeeman field $\Gamma$. We assume that the SM wire has disorder, which is described by the random potential $V_{dis}$, with $\langle V_{dis} \rangle=0$ and $\langle V_{dis}^2 \rangle = V_0^2$.  The specific disorder profile used in the numerical calculations is shown in Fig. \ref{FIG1}(b).
The self-energy describing the superconducting proximity effect is assumed to be local, with contributions of the form
\begin{equation}
[\Sigma_{SC}(\omega)]_{ii} = -\gamma \frac{\omega + \Delta_0 \sigma_y\tau_y}{\sqrt{\Delta_0^2-\omega^2}},
\end{equation}
where $\sigma_y$ and $\tau_y$ are Pauli matrices associated with the spin and particle-hole degrees of freedom, respectively, $\Delta_0$ is the gap of the parent superconductor (e.g., Al), and $\gamma=0.2~$meV is the effective SM-SC coupling. We assume that the SC gap decreases as a function of the applied field as shown in Fig. \ref{FIG1}(c), having a maximum value $\Delta_0(0)=0.3~$meV and completely collapsing at $\Gamma=1.25~$meV, which for typical values of the g-factor characterizing InAs wires, $g\sim 10-15$, corresponds to a magnetic field $B\sim 2.9-4.3~$T. 
Since we focus on the low-energy physics, e.g., on the Majorana modes that may emerge in the hybrid nanowire [see Fig. \ref{FIG1}(a)], it is convenient to work in the so-called static approximation, $\sqrt{\Delta_0^2-\omega^2}\approx \Delta_0(\Gamma)$, and define the effective BdG Hamiltonian
\begin{equation}
H_{eff} = Z\cdot H_{SM} + {\mathbb I}\cdot \Delta \sigma_y \tau_y, \label{Heff}
\end{equation}
where $Z=\Delta_0/(\Delta_0 +\gamma)$ is the quasiparticle residue, ${\mathbb I}_{ij} = \delta_{ij}$, and $\Delta=\Delta_0 ~\!\gamma/(\Delta_0 +\gamma)$ is the induced pairing potential. Note that the quantities $Z$ and $\Delta$ depend on the applied Zeeman field (through $\Delta_0$), decreasing from the maximum values $Z(0)=0.6$, $\Delta(0) = 0.12~$meV that correspond to zero magnetic field and eventually vanishing at $\Gamma=1.25~$meV. Also note that the ``bare'' parameters characterizing the SM Hamiltonian in Eq. (\ref{HSM}) are renormalized in the effective Hamiltonian by a factor $Z(\Gamma)$. 

As shown in Fig. \ref{FIG1}(a), the hybrid nanowire is coupled to a quantum dot (QD) characterized by a discrete set of states with energies $\epsilon_n+V_{QD}$ that can be tuned using a gate potential. We assume that the energy spacing is large compared with the relevant energy scales and, for simplicity, we focus on a single QD state, $n_0$, with $\epsilon_{n_0}=0$. The corresponding amplitudes of the left and right dot-nanowires couplings are $t\lambda_L$ and $t\lambda_R$, respectively, while the phases are controlled by the magnetic flux $\Phi$ threading the loop. Explicitly, the (second quantized) Hamiltonian describing the QD and the nanowire-QD coupling has the form
\begin{eqnarray}
&~& \hat{H}_{QD} = \sum_{\sigma}(V_{QD}-\mu)a_{\sigma}^\dagger a_{\sigma}^{} \label{HQD}\\
&-&t\sum_\sigma\left[(\lambda_L~\!e^{i\varphi} c_{1\sigma}^\dagger a_\sigma +{\rm h.c.}) + (\lambda_R ~\!e^{-i\varphi} c_{N\sigma}^\dagger a_\sigma +{\rm h.c.})\right], \nonumber 
\end{eqnarray}
where $a_{\sigma}^\dagger\equiv c_{0\sigma}^\dagger$ ($a_{\sigma}^{} \equiv c_{0\sigma}^{}$) is the creation (annihilation) operator corresponding to the dot state $n_0$ with spin $\sigma$ and $\varphi=(\pi/2)~\!\Phi/\Phi_0$, with $\Phi_0=h/2e$, is a flux-dependent phase. Here, the notation $c_{0\sigma}$ corresponds to labeling the QD as the site $i=0$ of the 1D lattice. 



\section{Results:} 
\subsection{Flux-dependent capacitance oscillation for topologically-protected Majorana modes with negligible overlap} 
\begin{figure}[t]
\begin{center}
\includegraphics[width=0.48\textwidth]{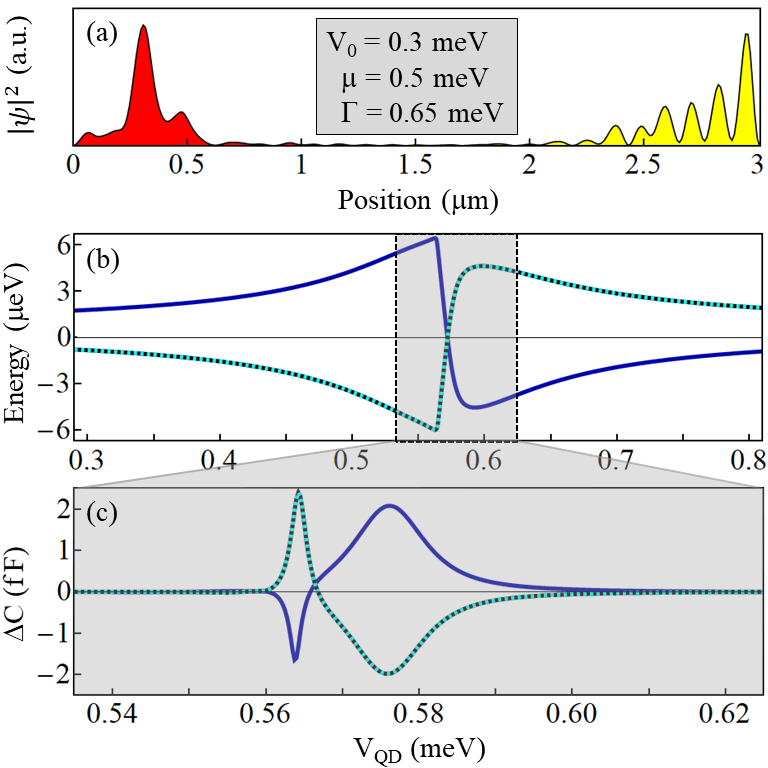}
\end{center}
\vspace{-2mm}
\caption{(a) Spatial profiles of Majorana modes supported by a three micron-long wire with low disorder of amplitude $V_0=0.3~$meV. The overlap of the two Majorana modes (see main text) is ${\cal O}=0.00056$. (b) Dependence of the lowest energy mode (corresponding to a fixed parity) on the quantum dot potential for $\Phi=0$ (dark blue) and $\Phi = h/2e$ (dashed line). The disorder and control parameters are the same as in panel (a) and the coupling to the quantum dot corresponds to $\lambda_L=\lambda_R=0.035$. Note the ``resonance'' (and the sign changes) corresponding to $V_{QD}\gtrsim \mu$. (c) Capacitance difference between different parity states as function of the quantum dot potential, $V_{QD}$, for $\Phi=0$ (dark blue) and $\Phi = h/2e$ (dashed line).}
\label{FIG2}
\vspace{0mm}
\end{figure}

To benchmark our calculations, we first consider a hybrid nanowire with low disorder ($V_0=0.3~$meV) and tune the control parameters to a specific point within the topological phase corresponding to $\mu=0.5~$meV and $\Gamma=0.65~$meV. The lowest energy state of the effective BdG Hamiltonian given by Eq. (\ref{Heff}) corresponds to the pair of Majorana modes with spatial profiles given in Fig. \ref{FIG2}(a). To characterize the spatial separation of the Majorana modes we define the ``overlap'' of the corresponding probability distributions,
\begin{equation}
{\cal O} =\frac{\sum_\alpha |\psi_{M_1}(\alpha)|^2|\psi_{M_2}(\alpha)|^2}{\sqrt{\sum_\alpha |\psi_{M_1}(\alpha)|^4\sum_\alpha |\psi_{M_2}(\alpha)|^4}},
\end{equation}
where $\alpha=(i,\sigma,\tau)$, representing position, spin, and particle-hole labels, takes $4N$ values and $\psi_{M_\ell}(\alpha)$ is the wave function of the $\ell$ Majorana mode. Note that ${\cal O}=1$ corresponds to a pair of perfectly overlapping Majoranas, while ${\cal O}=0$ indicates a pair of perfectly separated Majorana modes. The example shown in Fig. \ref{FIG2}(a) corresponds to a pair of well separated topologically protected Majoranas with ${\cal O}=5.6\times 10^{-4}$.  

Next, we couple the hybrid wire to the quantum dot and diagonalize the corresponding BdG Hamiltonian, $H_{eff}+H_{QD}$. The low-energy states of the coupled system depend on the wire-QD couplings, $\lambda_L$ and $\lambda_R$, the gate potential on the dot, $V_{QD}$, and the flux $\Phi$ that threads the interferometric loop. As an example, in Fig. \ref{FIG2}(b) we show the dependence of the lowest energy BdG mode on the QD potential, $V_{QD}$, for a system with $\lambda_L=\lambda_R=0.035$ and magnetic flux $\Phi=0$ (continuous dark blue line) or $\Phi=h/2e$ (dashed line). Note that the two curves correspond to a state with well defined (odd) parity; the state with opposite (even) parity has energy $E_e(V_{QD},\Phi)=-E_o(V_{QD},\Phi)$. The strongest dependence on the dot potential corresponds to a ``resonance'' near $V_{QD}$ values slightly larger than the chemical potential. Within the resonance region, we calculate the zero frequency capacitance for the odd and even parity sectors, which is given by the $\omega=0$ value of the dynamical capacitance \cite{Microsoft2025,Sau2024}
\begin{equation}
C(\omega) =\kappa \sum_{n, m}^* \frac{|\langle \psi_n|\tau_z ~\!\delta_{i 0}|\psi_m\rangle|^2(E_n-E_m)}{(E_n-E_m)^2-(\omega+i\eta)^2}, \label{Cw} 
\end{equation}
where $\kappa = 0.032~$fF$\cdot~\mu$eV and $\eta=2~\mu$eV. The matrix elements in the numerator are calculated using the BdG states $|\psi_n\rangle$ and $|\psi_m \rangle$ of energies $E_n$ and $E_m$, respectively, and the summation corresponding to the odd (even) sector is constrained as follows: $n\geq 2N+2$ and $n=e~\!(o)$, $m\leq 2N-1$ and $m=o~\!(e)$, with $e$ ($o$) being one of the lowest energy modes, $2N$ or $2N+1$, of energies $E_e=-E_o$. The presence of a constraint is indicated by the ``star'' symbol in Eq. (\ref{Cw}). The capacitance difference between the even and odd sector, $\Delta C = C_e - C_o$, as function of the quantum dot potential, $V_{QD}$, is shown in Fig. \ref{FIG2}(c) for a system with magnetic flux through the interferometric loop $\Phi=0$ (continuous dark blue line) and $\Phi=h/2e$ (dashed line). Note that $\Delta C$ has significant values (on the order of $1~$fF) within a potential window of about $0.04~$meV --- the ``resonance'' region corresponding to $V_{QD}$ values slightly above the chemical potential. Also note that $\Delta C$ changes sign at $V_{QD}\approx 0.567~$meV, indicating a switch of the capacitance magnitudes characterizing the even and odd parity sectors. Finally, the signs of $\Delta C$ corresponding to $\Phi=0$ and $\Phi=h/2e$ are typically opposite, which suggests that the capacitance values characterizing different parity sectors oscillate as function of the applied magnetic flux.     

\begin{figure}[t]
\begin{center}
\includegraphics[width=0.48\textwidth]{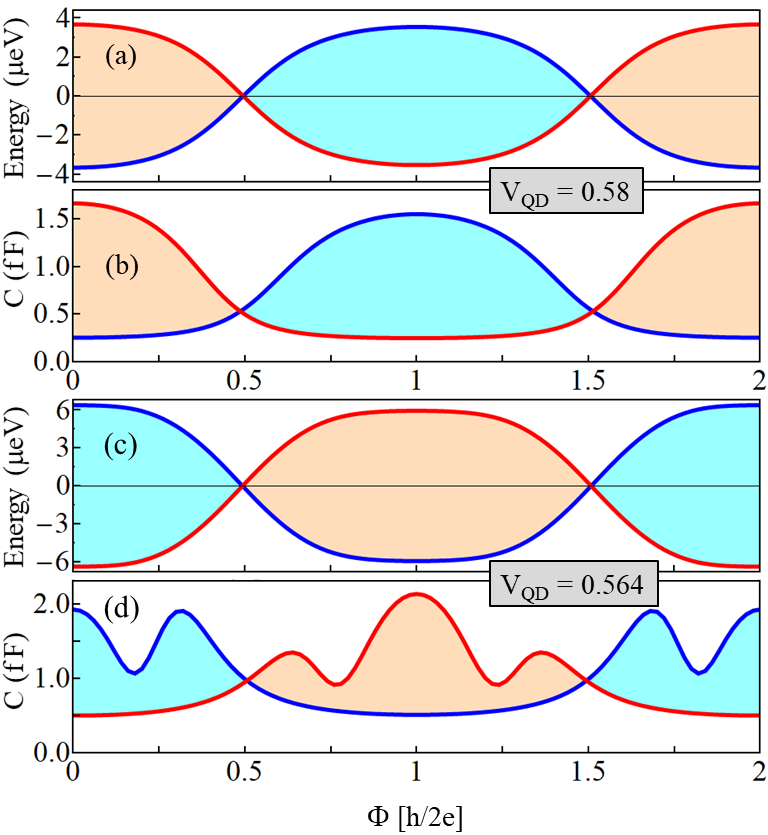}
\end{center}
\vspace{-2mm}
\caption{Flux-induced oscillations of the energy and capacitance corresponding to different parity states of a (low-disorder) nanowire-quantum dot system with parameters as in Fig. \ref{FIG2}. (a) Energy oscillations and (b) capacitance corresponding to a dot potential $V_{QD}$ (given in meV) within the right ``resonance lobe'' in Fig. \ref{FIG2}(c). (c) Energy and (d) capacitance as functions of the magnetic flux $\Phi$ for a dot potential within the left ``resonance lobe'' in Fig. \ref{FIG2}(c).}
\label{FIG3}
\vspace{0mm}
\end{figure}

The flux-induced oscillations of the capacitance and of the corresponding lowest energy modes are illustrated in Fig. \ref{FIG3}. We consider two values of the dot potential, one within the right ``resonance lobe'' ($V_{QD}=0.58~$meV) and the other within the left ``lobe'' ($V_{QD}=0.564~$meV) --- also see Fig. \ref{FIG2}(c). For a given parity, the period of the oscillations is $h/e$. The amplitude of the energy oscillations, which is on the order of a few $\mu$eV, depends strongly on the coupling between the wire and the quantum dot, i.e., on the parameters $\lambda_L$ and $\lambda_R$. The capacitance oscillations have the same main features as the energy oscillations, but can exhibit specific details, like, e.g., the presence of a higher oscillatory component in Fig. \ref{FIG3}(d). The key point is that in the presence of weak disorder the hybrid Majorana wire supports well separated Majorana modes [Fig. \ref{FIG2}(a)], which in an interferometric device [Fig. \ref{FIG1}(a)] lead to flux-induced oscillations of the capacitance corresponding to the odd and even parity sectors. The period of the oscillations (for a given parity) is $h/e$ and the maximum difference between the capacitance values corresponding to the two sectors is on the order of $1~$fF.  

\begin{figure}[t]
\begin{center}
\includegraphics[width=0.48\textwidth]{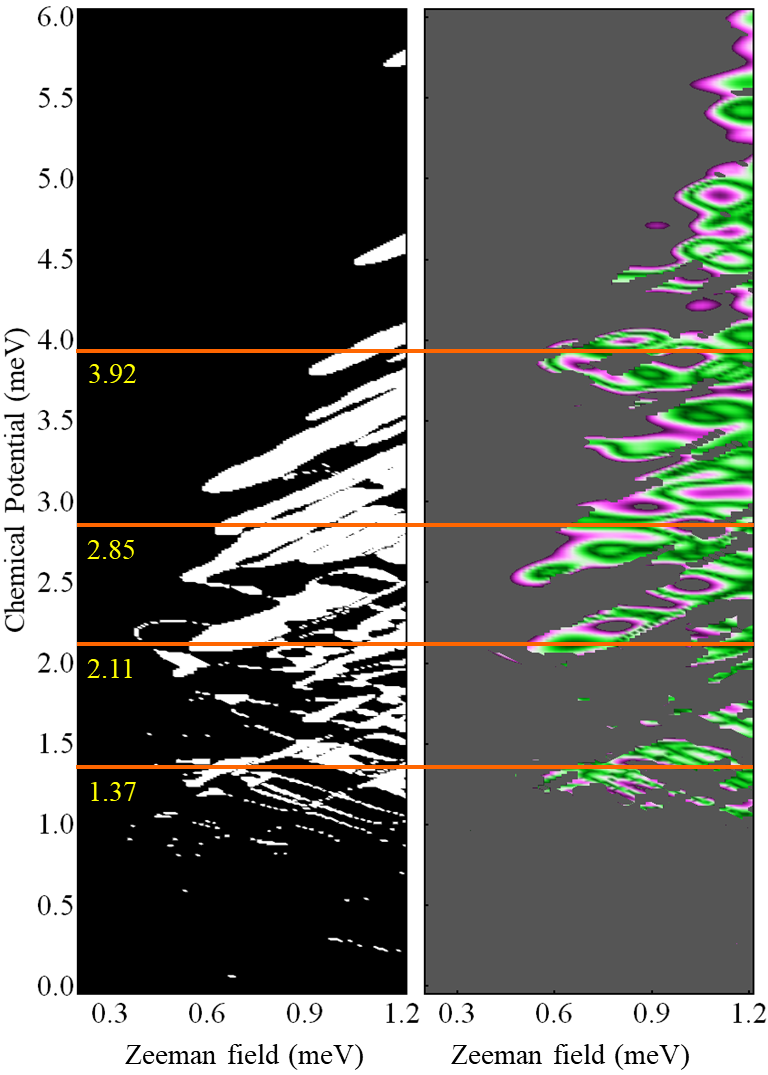}
\end{center}
\vspace{-2mm}
\caption{{\em Left}: Topological phase diagram for a hybrid nanowire with strong disorder ($V_0=1.2~$meV). The white and black regions are topologically non-trivial and trivial, respectively. The topological invariant was defined for an infinite system obtained by periodically repeating the $L=3~\mu$m disorder wire (see main text). {\em Right}: Low-energy map showing the control parameter region that supports delocalized states (see main text) with energies between zero (dark green) and $10~\mu$eV (pink). Parameter values within the dark gray region correspond to lowest energy states that are either localized, or have energies larger than $10~\mu$eV. Note the significant correlation between the left and right panels for chemical potential values below $\mu \approx 3~$meV. The low-energy spectra along the horizontal cuts marked by orange lines are shown in Fig. \ref{FIG5}.}
\label{FIG4}
\vspace{0mm}
\end{figure}

\subsection{Flux-dependent capacitance oscillation for partially-separated Majorana modes and quasi-Majorana modes} What is the corresponding picture in a system with intermediate-to-strong disorder? To address this question, we consider a system with the same parameters as those used above, except the disorder amplitude, which is now $V_0=1.2~$meV. We first identify the control parameter regimes consistent with the presence of Majorana physics, e.g., ``patches'' within which the system could be considered as being in some operationally-defined ``topological'' phase. The topological map calculated using the winding number invariant for the infinite system obtained by periodically repeating the finite disordered wire \cite{Eissele2025} is shown in Fig. \ref{FIG4} (left panel). Note that the topological region (white) is characterized by chemical potential values significantly larger than the $-1.2 \lesssim \mu \lesssim 1.2~$meV window corresponding to the clean system. In addition, the topological region is highly fragmented, indicating the presence of large finite size effects \cite{Eissele2025}. Relatively stable, nearly-zero energy Majorana modes are expected to emerge within the large topological ``islands.'' We also map the lowest energy of the (finite) wire as function of the Zeeman field and chemical potential, under the constraint that the corresponding mode be delocalized. Specifically, we require that $90\%$ of the spectral weight of the lowest energy mode be distributed over at least $85\%$ of the wire length. The corresponding map is shown in the right panel of Fig. \ref{FIG4}, with dark gray corresponding to either localized (lowest energy) states, or (gapped) states with energies larger than $10~\mu$eV. The dark green regions indicate the presence of nearly-zero energy (delocalized) modes. Note the significant correlation between the topological map (left) and the low-energy map (right) for chemical potential values $\mu \lesssim 3~$meV.        

\begin{figure}[t]
\begin{center}
\includegraphics[width=0.48\textwidth]{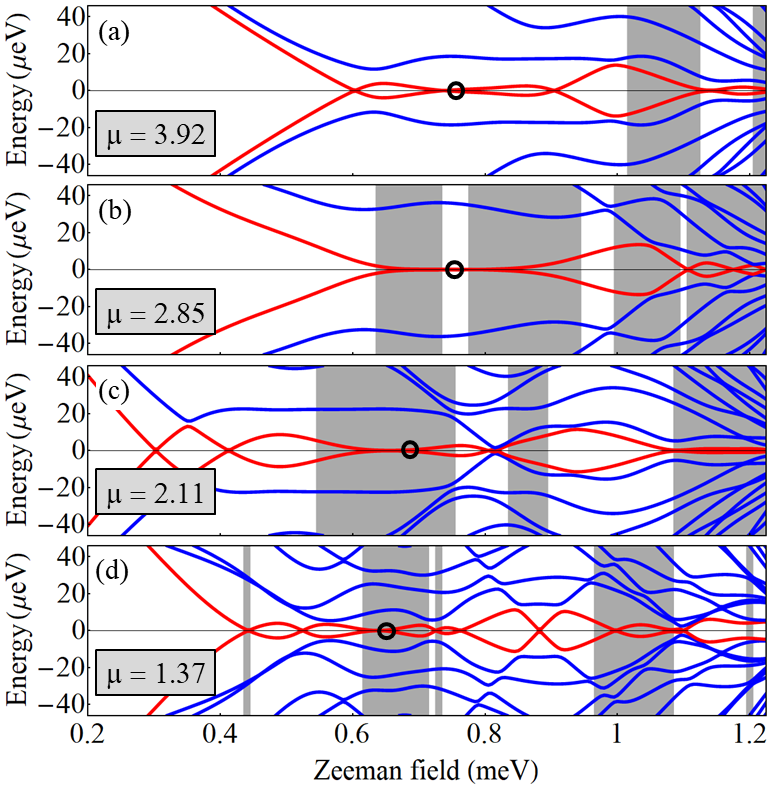}
\end{center}
\vspace{-2mm}
\caption{Low-energy spectra as functions of the applied Zeeman field for a system with strong disorder ($V_0=1.2~$meV) and different values of the chemical potential (given in meV) corresponding to the cuts marked in Fig. \ref{FIG4}. The shaded regions correspond to a nontrivial value of the topological invariant. The spatial profiles of the Majorana modes corresponding to the nearly-zero energy states marked by the black circles are shown in Fig. \ref{FIG6}.}
\label{FIG5}
\vspace{0mm}
\end{figure}

We now select four constant-$\mu$ cuts through parameter regions that support nearly-zero energy (delocalized) modes, as indicated by the orange lines in Fig. \ref{FIG4}. For $\mu\lesssim 3~$meV those regions correspond to topologically-nontrivial patches (see Fig. \ref{FIG4}), while the cut at $\mu=3.92~$meV includes a relatively large window ($0.6\lesssim \mu\lesssim 0.9~$meV) that supports a low-energy topologically-trivial mode, whose constituent Majorana wave functions are slightly displaced from each other, making it a partially-seperated ABS or quasi-Majorana mode (q-MZM). The corresponding energy spectra (as function of the applied Zeeman field) are shown in Fig. \ref{FIG5}. As expected \cite{Eissele2025}, the (large) topological regions (gray shading) support relatively robust low-energy modes, with typical energy splittings of up to a few $\mu$eV. Of course, enhancing the robustness of these modes would require longer Majorana wires \cite{Eissele2025}. In addition, in Fig. \ref{FIG5}(a) we show an example of a (relatively robust) topologically trivial low-energy mode which is a quasi-Majorana mode. Note that the nearly-zero energy modes hosted by the low-field topological regions in panels (b) and (c) are protected by energy gaps of about $30~\mu$eV and $20~\mu$eV, respectively, and that there is no signature corresponding to the closing of this gap at a topological quantum phase transition. This behavior indicates the presence of significant finite-size effects \cite{Eissele2025} and that these regions of parameter space do not fulfill the topological gap protocol \cite{PhysRevB.107.245423} 
employed to select experimental parameters for observing flux-dependent capacitance oscillations. Remarkably, we find that the systems shown in panels (b) and (c) exhibit flux-dependent oscillations in quantum capacitance similar to the experiments despite not showing closing and reopening of bulk gap and therefore not satisfying the topological gap protocol, and the system in panel (a) does so as well, even though it is topologically trivial.

\begin{figure}[t]
\begin{center}
\includegraphics[width=0.48\textwidth]{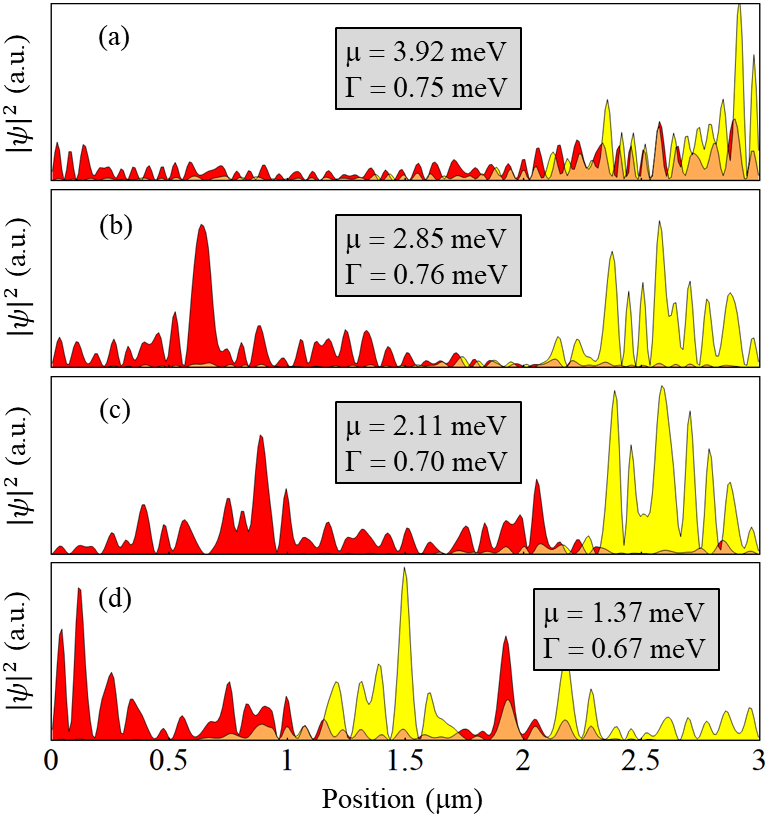}
\end{center}
\vspace{-2mm}
\caption{Spatial profiles of the Majorana modes associated with the near-zero-energy states marked in Fig. \ref{FIG5}. The control parameters are explicitly given in the corresponding panels. Note that all these states can be viewed as partially separated (or partially overlapping) Majorana modes. The corresponding overlaps are: (a) ${\cal O}=0.63$; (b) ${\cal O}=0.046$; (c) ${\cal O}=0.066$; (d) ${\cal O}=0.23$. The lowest overlap, corresponding to the modes in ((b), is almost two orders of magnitude larger than the overlap characterizing the low disorder modes in Fig. \ref{FIG2}(a).}
\label{FIG6}
\vspace{0mm}
\end{figure}

The final step of the selection process involves choosing specific (nearly) zero energy states. We focus on the low-field regime ($\Gamma \lesssim 0.8~$meV), which is more accessible, hence more relevant experimentally. The selected states are marked by black circles in Fig. \ref{FIG5}, while the spatial profiles of the corresponding Majorana 
\begin{figure}[t]
\begin{center}
\includegraphics[width=0.48\textwidth]{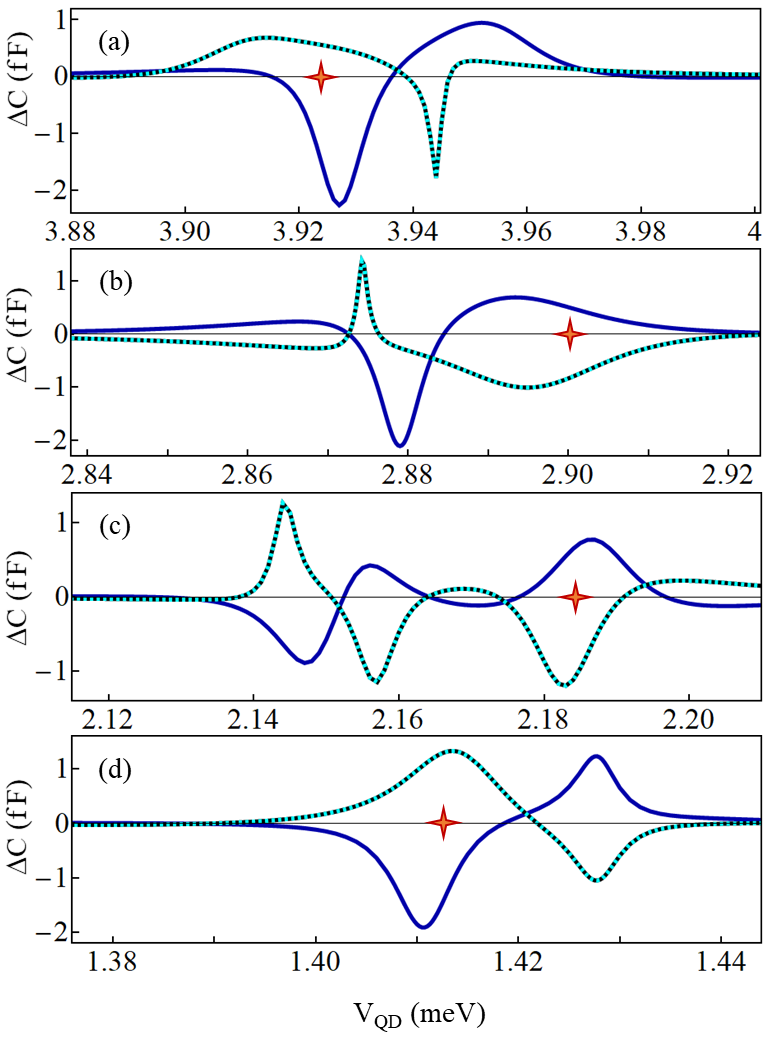}
\end{center}
\vspace{-2mm}
\caption{Capacitance difference between different parity states as function of the quantum dot potential, $V_{QD}$, for a nanowire-quantum dot system with control parameters given in the corresponding panels of Fig. \ref{FIG6}. The dark blue lines are for a system with $\Phi=0$, while the dashed lines correspond to $\Phi=h/2e$. The flux-dependent oscillations of the capacitance corresponding to the dot potential values marked by red crosses are shown in Fig. \ref{FIG8}. The nanowire-quantum dot coupling parameters and the specific $V_{QD}$ values corresponding to the red crosses are: (a) $\lambda_L=0.02$, $\lambda_R=0.01$, $V_{QD}=3.922~$meV; (b) $\lambda_L=0.025$, $\lambda_R=0.025$, $V_{QD}=2.9~$meV; (c) $\lambda_L=0.035$, $\lambda_R=0.03$, $V_{QD}=2.185~$meV; (d) $\lambda_L=0.025$, $\lambda_R=0.025$, $V_{QD}=1.413~$meV.} 
\label{FIG7}
\vspace{-2mm}
\end{figure}
modes are shown in Fig. \ref{FIG6}.  The low-energy states correspond to pairs of partially separated Majorana modes with overlaps ranging from ${\cal O}=0.046$ [panel (b)] to quasi-Majorana modes with ${\cal O}=0.63$ [panel (a)]. Note that the lowest overlap (${\cal O}=0.046$) is about two orders of magnitude larger than the overlap characterizing the well-separated Majorana modes shown in Fig. \ref{FIG2}(a) and the overlaps in panels (c) and (d) are significantly larger than in panel (b). This property is generic and is due to the fact that in a (strongly) disordered system the low-energy Majorana modes are not exponentially localized near the ends of the wire, but have relatively large characteristic length scales ranging from a few microns up to tens of microns, depending on the specific system parameters (e.g., SM-SC coupling, disorder amplitude, etc.). In general, the values of the characteristic length scale increase with increasing the disorder strength (as the topological phase ``shifts'' toward larger values of the chemical potential and Zeeman field \cite{Roy_2024,Eissele2025}) or with reducing the effective SM-SC coupling (which implies lower values of the renormalized effective mass, Fermi velocity, etc.). We point out that the system also supports strongly localized low-energy states (typically at low values of the chemical potential), but these states are topologically trivial (see the topological map in Fig. \ref{FIG4}). Thus, the generic delocalized low-energy states emerging in (strongly) disordered Majorana hybrid wires of lengths on the order of a few microns consist of partially separated Majorana modes (Fig. 6, panels (b,c,d)) characterized by typical overlap values significantly larger than those associated with their clean system counterparts (Fig. 2, panel (a)) or topologically trivial quasi-Majorana modes (Fig. 6, panel (a)). We emphasize that labeling some of these modes, e.g., those shown in Fig. \ref{FIG6}(b) and (c), as ``topological'' Majorana zero modes is a matter of taste (i.e., dependent on the operational definition of ``topology''), but, in general, there exist (relatively robust) low-energy q-MZMs that are definitely topologically trivial, e.g., the state shown in Fig. \ref{FIG6}(a).         


\begin{figure}[t]
\begin{center}
\includegraphics[width=0.48\textwidth]{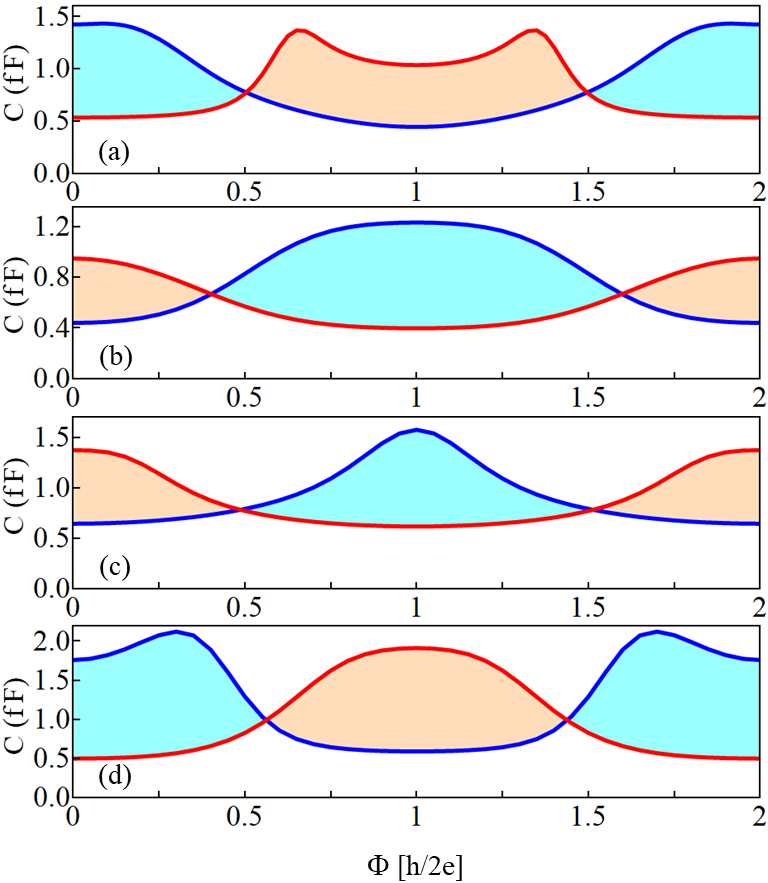}
\end{center}
\vspace{-2mm}
\caption{Flux-dependent oscillations of the capacitance corresponding to the control parameter values given in Fig. \ref{FIG6} and the quantum dot potential values indicated in Fig. \ref{FIG7}. Note that the main features, and even some quantitative aspects (e.g., the amplitude of the oscillations), are basically independent of the underlying low-energy states and are similar to those characterizing the ``ideal'' case shown in Fig. \ref{FIG3}.}
\label{FIG8}
\vspace{0mm}
\end{figure}

Having identified a set of representative (delocalized) nearly-zero energy modes hosted by the Majorana hybrid wire with intermediate-to-strong disorder, we calculate the quantum capacitance signatures associated with different parity sectors using the interferometric setup consisting of the Majorana wire coupled to a quantum dot (see Fig. \ref{FIG1}). The dependence of the capacitance difference between the even and odd parity sectors on the quantum dot potential is shown in Fig. \ref{FIG7}. Panels (a)--(d) show $\Delta C$ for a system with magnetic flux $\Phi=0$ (dark blue) and $\Phi=h/2e$ (dashed lines) and control parameters (i.e., Zeeman field and chemical potential) given in the corresponding panels of Fig. \ref{FIG6}. Note that $\Delta C$ has significant values (on the order of $1~$fF) within narrow ``resonance'' regions corresponding to $V_{QD}$ values slightly above the chemical potential and changes sign at least once, qualitatively similar to the ``ideal'' case illustrated in Fig. \ref{FIG2}. 

Finally, we calculate the dependence of the capacitance characterizing the odd and even parity sectors on the magnetic flux $\Phi$ for specific values of the quantum dot potential  marked by red crosses in Fig. \ref{FIG7}. The parity-dependent oscillations shown in the corresponding panels of Fig. \ref{FIG8} represent the main result of this work. For a given parity, the period of these oscillations is $h/e$, while their amplitude is on the order
of $1~$fF, similar to the ``ideal'' (low-disorder) scenario illustrated in Fig. \ref{FIG3}. We emphasize that there are no qualitative (or even significant quantitative) differences among the four panels in Fig. \ref{FIG8}, or between these cases and the ''ideal'' scenario shown in Fig. \ref{FIG3}. In other words, parity-selective oscillations of the capacitance (having period $h/e$) measured in an interferometric device consisting of a disordered, few micron-long Majorana wire coupled to a quantum dot can indicate the presence of partially-separated and strongly overlapping Majorana modes and even topologically trivial quasi-Majorana modes \cite{Moore2018,moore2018two,vuik2018reproducing}. However, these modes are not necessarily topologically protected (i.e., well separated) and may even emerge in a topologically trivial phase. The probability that the partially-separated Majorana modes responsible for the conductance oscillations be topologically protected decreases strongly with the disorder strength.   

\section{Conclusion} Motivated by the interesting results of a recent experiment \cite{Microsoft2025}, we consider an interferometric device consisting of a hybrid semiconductor-superconductor nanowire coupled to a quantum dot and calculate the quantum capacitance of the dot as a function of the magnetic flux through the interferometric loop for control parameter values corresponding to representative low-energy states of the wire. The experiment has reported 
a fermion parity-dependent shift of the quantum capacitance of up to $1~$fF and the observation of flux $h/2e$-periodic bimodality, which is consistent with parity-selective capacitance oscillations of period $h/e$. We model the semiconductor wire using a simple tight-binding Hamiltonian and incorporate the proximity effects induced by the parent superconductor through a self-energy term. The effective model parameters are consistent with the values characterizing the InAs–Al hybrid devices used in the experiment. 

After benchmarking our quantum capacitance calculations using a low-disorder system that hosts a pair of nearly ``ideal'' pair of Majorana zero modes, we focus on a hybrid nanowire with intermediate-to-strong disorder. First, we map the relevant control parameter space and show that the generic delocalized low-energy states supported by the $3$ micron-long hybrid nanowire consist of partially-separated Majorana modes. The characteristic length scales of these modes are on the order of $1-3$ microns, comparable to the length of the system, and their overlap is at least two orders of magnitude larger than the overlap characterizing the ``ideal'' pair of Majoranas hosted by the low-disorder wire. Furthermore, while some of these Majorana modes could be dubbed as being ``topological'' (within a relaxed operational definition of ``topology''), others are clearly topologically trivial quasi-Majorana modes. We emphasize that, for the given values of the system parameters, enhancing the likelihood of realizing ``topological'' Majorana modes (i.e., well-separated modes with low overlap values) requires reducing the disorder strength. However, this would lead to the emergence of large topological regions (in the control parameter space), which appears to be inconsistent with the experimental situation. On the other hand, increasing the disorder strength would enhance the fragmentation of the ``topological'' phase and would increase the (typical) values of the overlap and the likelihood of having (relatively robust) nearly-zero trivial states.

The main result of our work, which is illustrated in Fig. \ref{FIG8}, concerns the emergence of parity-dependent oscillations of the quantum capacitance of the dot as a function of the magnetic flux threading the interferometric loop. These oscillations, which have period $h/e$ and amplitudes on the order of $1~$fF (consistent with the recent experimental observations), are associated with nearly zero energy states emerging in the nanowire that, in general, consist of a pair of partially overlapping Majorana modes or even a topologically trivial quasi-Majorana mode. The corresponding overlap values, which can be as high as $65\%$, do not affect the main features of the parity-dependent oscillations. We conclude that the observation of such oscillations indicates the presence of partially separated Majorana modes or topologically trivial quasi-Majorana modes and does not represent a unique signature of topologically-protected Majorana zero modes. 

\section{Acknowledgment}
T.D.S. thanks ONR-N000142312061 for support. S.T. acknowledges support from SC Quantum, ARO W911NF2210247, and ONR-N000142312061. S. T. thanks Jay D. Sau for fruitful discussions. 

\bibliography{REF}

\begin{thebibliography}{56}%
\makeatletter
\providecommand \@ifxundefined [1]{%
 \@ifx{#1\undefined}
}%
\providecommand \@ifnum [1]{%
 \ifnum #1\expandafter \@firstoftwo
 \else \expandafter \@secondoftwo
 \fi
}%
\providecommand \@ifx [1]{%
 \ifx #1\expandafter \@firstoftwo
 \else \expandafter \@secondoftwo
 \fi
}%
\providecommand \natexlab [1]{#1}%
\providecommand \enquote  [1]{``#1''}%
\providecommand \bibnamefont  [1]{#1}%
\providecommand \bibfnamefont [1]{#1}%
\providecommand \citenamefont [1]{#1}%
\providecommand \href@noop [0]{\@secondoftwo}%
\providecommand \href [0]{\begingroup \@sanitize@url \@href}%
\providecommand \@href[1]{\@@startlink{#1}\@@href}%
\providecommand \@@href[1]{\endgroup#1\@@endlink}%
\providecommand \@sanitize@url [0]{\catcode `\\12\catcode `\$12\catcode `\&12\catcode `\#12\catcode `\^12\catcode `\_12\catcode `\%12\relax}%
\providecommand \@@startlink[1]{}%
\providecommand \@@endlink[0]{}%
\providecommand \url  [0]{\begingroup\@sanitize@url \@url }%
\providecommand \@url [1]{\endgroup\@href {#1}{\urlprefix }}%
\providecommand \urlprefix  [0]{URL }%
\providecommand \Eprint [0]{\href }%
\providecommand \doibase [0]{https://doi.org/}%
\providecommand \selectlanguage [0]{\@gobble}%
\providecommand \bibinfo  [0]{\@secondoftwo}%
\providecommand \bibfield  [0]{\@secondoftwo}%
\providecommand \translation [1]{[#1]}%
\providecommand \BibitemOpen [0]{}%
\providecommand \bibitemStop [0]{}%
\providecommand \bibitemNoStop [0]{.\EOS\space}%
\providecommand \EOS [0]{\spacefactor3000\relax}%
\providecommand \BibitemShut  [1]{\csname bibitem#1\endcsname}%
\let\auto@bib@innerbib\@empty
\bibitem [{\citenamefont {Read}\ and\ \citenamefont {Green}(2000)}]{Read2000}%
  \BibitemOpen
  \bibfield  {author} {\bibinfo {author} {\bibfnamefont {N.}~\bibnamefont {Read}}\ and\ \bibinfo {author} {\bibfnamefont {D.}~\bibnamefont {Green}},\ }\bibfield  {title} {\bibinfo {title} {Paired states of fermions in two dimensions with breaking of parity and time-reversal symmetries and the fractional quantum hall effect},\ }\href {https://doi.org/10.1103/PhysRevB.61.10267} {\bibfield  {journal} {\bibinfo  {journal} {Phys. Rev. B}\ }\textbf {\bibinfo {volume} {61}},\ \bibinfo {pages} {10267} (\bibinfo {year} {2000})}\BibitemShut {NoStop}%
\bibitem [{\citenamefont {Kitaev}(2001)}]{Kitaev_2001}%
  \BibitemOpen
  \bibfield  {author} {\bibinfo {author} {\bibfnamefont {A.~Y.}\ \bibnamefont {Kitaev}},\ }\bibfield  {title} {\bibinfo {title} {Unpaired majorana fermions in quantum wires},\ }\href {https://doi.org/10.1070/1063-7869/44/10S/S29} {\bibfield  {journal} {\bibinfo  {journal} {Physics-Uspekhi}\ }\textbf {\bibinfo {volume} {44}},\ \bibinfo {pages} {131} (\bibinfo {year} {2001})}\BibitemShut {NoStop}%
\bibitem [{\citenamefont {Kitaev}(2003)}]{Kitaev2003}%
  \BibitemOpen
  \bibfield  {author} {\bibinfo {author} {\bibfnamefont {A.~Y.}\ \bibnamefont {Kitaev}},\ }\href@noop {} {\bibfield  {journal} {\bibinfo  {journal} {Annals. Phys.}\ }\textbf {\bibinfo {volume} {303}},\ \bibinfo {pages} {2} (\bibinfo {year} {2003})}\BibitemShut {NoStop}%
\bibitem [{\citenamefont {Majorana}(1937)}]{Majorana1937}%
  \BibitemOpen
  \bibfield  {author} {\bibinfo {author} {\bibfnamefont {E.}~\bibnamefont {Majorana}},\ }\href@noop {} {\bibfield  {journal} {\bibinfo  {journal} {Nuovo Cimento}\ }\textbf {\bibinfo {volume} {14}},\ \bibinfo {pages} {171} (\bibinfo {year} {1937})}\BibitemShut {NoStop}%
\bibitem [{\citenamefont {Wilczek}(1982)}]{wilczek1982quantum}%
  \BibitemOpen
  \bibfield  {author} {\bibinfo {author} {\bibfnamefont {F.}~\bibnamefont {Wilczek}},\ }\bibfield  {title} {\bibinfo {title} {Quantum mechanics of fractional-spin particles},\ }\href@noop {} {\bibfield  {journal} {\bibinfo  {journal} {Physical review letters}\ }\textbf {\bibinfo {volume} {49}},\ \bibinfo {pages} {957} (\bibinfo {year} {1982})}\BibitemShut {NoStop}%
\bibitem [{\citenamefont {Moore}\ and\ \citenamefont {Read}(1991)}]{Moore1991}%
  \BibitemOpen
  \bibfield  {author} {\bibinfo {author} {\bibfnamefont {G.}~\bibnamefont {Moore}}\ and\ \bibinfo {author} {\bibfnamefont {N.}~\bibnamefont {Read}},\ }\href@noop {} {\bibfield  {journal} {\bibinfo  {journal} {Nucl. Physics B}\ }\textbf {\bibinfo {volume} {360}},\ \bibinfo {pages} {362} (\bibinfo {year} {1991})}\BibitemShut {NoStop}%
\bibitem [{\citenamefont {Nayak}\ and\ \citenamefont {Wilczek}(1996)}]{Nayak1996}%
  \BibitemOpen
  \bibfield  {author} {\bibinfo {author} {\bibfnamefont {C.}~\bibnamefont {Nayak}}\ and\ \bibinfo {author} {\bibfnamefont {F.}~\bibnamefont {Wilczek}},\ }\href@noop {} {\bibfield  {journal} {\bibinfo  {journal} {Nucl. Physics B}\ }\textbf {\bibinfo {volume} {479}},\ \bibinfo {pages} {529} (\bibinfo {year} {1996})}\BibitemShut {NoStop}%
\bibitem [{\citenamefont {Nayak}\ \emph {et~al.}(2008)\citenamefont {Nayak}, \citenamefont {Simon}, \citenamefont {Stern}, \citenamefont {Freedman},\ and\ \citenamefont {Das~Sarma}}]{Nayak2008}%
  \BibitemOpen
  \bibfield  {author} {\bibinfo {author} {\bibfnamefont {C.}~\bibnamefont {Nayak}}, \bibinfo {author} {\bibfnamefont {S.~H.}\ \bibnamefont {Simon}}, \bibinfo {author} {\bibfnamefont {A.}~\bibnamefont {Stern}}, \bibinfo {author} {\bibfnamefont {M.}~\bibnamefont {Freedman}},\ and\ \bibinfo {author} {\bibfnamefont {S.}~\bibnamefont {Das~Sarma}},\ }\bibfield  {title} {\bibinfo {title} {Non-abelian anyons and topological quantum computation},\ }\href {https://doi.org/10.1103/RevModPhys.80.1083} {\bibfield  {journal} {\bibinfo  {journal} {Rev. Mod. Phys.}\ }\textbf {\bibinfo {volume} {80}},\ \bibinfo {pages} {1083} (\bibinfo {year} {2008})}\BibitemShut {NoStop}%
\bibitem [{\citenamefont {Sau}\ \emph {et~al.}(2010{\natexlab{a}})\citenamefont {Sau}, \citenamefont {Lutchyn}, \citenamefont {Tewari},\ and\ \citenamefont {Sarma}}]{sau2010generic}%
  \BibitemOpen
  \bibfield  {author} {\bibinfo {author} {\bibfnamefont {J.~D.}\ \bibnamefont {Sau}}, \bibinfo {author} {\bibfnamefont {R.~M.}\ \bibnamefont {Lutchyn}}, \bibinfo {author} {\bibfnamefont {S.}~\bibnamefont {Tewari}},\ and\ \bibinfo {author} {\bibfnamefont {S.~D.}\ \bibnamefont {Sarma}},\ }\bibfield  {title} {\bibinfo {title} {Generic new platform for topological quantum computation using semiconductor heterostructures},\ }\href@noop {} {\bibfield  {journal} {\bibinfo  {journal} {Physical review letters}\ }\textbf {\bibinfo {volume} {104}},\ \bibinfo {pages} {040502} (\bibinfo {year} {2010}{\natexlab{a}})}\BibitemShut {NoStop}%
\bibitem [{\citenamefont {Sau}\ \emph {et~al.}(2010{\natexlab{b}})\citenamefont {Sau}, \citenamefont {Tewari}, \citenamefont {Lutchyn}, \citenamefont {Stanescu},\ and\ \citenamefont {Sarma}}]{sau2010non}%
  \BibitemOpen
  \bibfield  {author} {\bibinfo {author} {\bibfnamefont {J.~D.}\ \bibnamefont {Sau}}, \bibinfo {author} {\bibfnamefont {S.}~\bibnamefont {Tewari}}, \bibinfo {author} {\bibfnamefont {R.~M.}\ \bibnamefont {Lutchyn}}, \bibinfo {author} {\bibfnamefont {T.~D.}\ \bibnamefont {Stanescu}},\ and\ \bibinfo {author} {\bibfnamefont {S.~D.}\ \bibnamefont {Sarma}},\ }\bibfield  {title} {\bibinfo {title} {Non-abelian quantum order in spin-orbit-coupled semiconductors: Search for topological majorana particles in solid-state systems},\ }\href@noop {} {\bibfield  {journal} {\bibinfo  {journal} {Physical Review B}\ }\textbf {\bibinfo {volume} {82}},\ \bibinfo {pages} {214509} (\bibinfo {year} {2010}{\natexlab{b}})}\BibitemShut {NoStop}%
\bibitem [{\citenamefont {Oreg}\ \emph {et~al.}(2010)\citenamefont {Oreg}, \citenamefont {Refael},\ and\ \citenamefont {von Oppen}}]{oreg2010helical}%
  \BibitemOpen
  \bibfield  {author} {\bibinfo {author} {\bibfnamefont {Y.}~\bibnamefont {Oreg}}, \bibinfo {author} {\bibfnamefont {G.}~\bibnamefont {Refael}},\ and\ \bibinfo {author} {\bibfnamefont {F.}~\bibnamefont {von Oppen}},\ }\bibfield  {title} {\bibinfo {title} {Helical liquids and majorana bound states in quantum wires},\ }\href@noop {} {\bibfield  {journal} {\bibinfo  {journal} {Physical review letters}\ }\textbf {\bibinfo {volume} {105}},\ \bibinfo {pages} {177002} (\bibinfo {year} {2010})}\BibitemShut {NoStop}%
\bibitem [{\citenamefont {Lutchyn}\ \emph {et~al.}(2010)\citenamefont {Lutchyn}, \citenamefont {Sau},\ and\ \citenamefont {Sarma}}]{lutchyn2010majorana}%
  \BibitemOpen
  \bibfield  {author} {\bibinfo {author} {\bibfnamefont {R.~M.}\ \bibnamefont {Lutchyn}}, \bibinfo {author} {\bibfnamefont {J.~D.}\ \bibnamefont {Sau}},\ and\ \bibinfo {author} {\bibfnamefont {S.~D.}\ \bibnamefont {Sarma}},\ }\bibfield  {title} {\bibinfo {title} {Majorana fermions and a topological phase transition in semiconductor-superconductor heterostructures},\ }\href@noop {} {\bibfield  {journal} {\bibinfo  {journal} {Physical review letters}\ }\textbf {\bibinfo {volume} {105}},\ \bibinfo {pages} {077001} (\bibinfo {year} {2010})}\BibitemShut {NoStop}%
\bibitem [{\citenamefont {Mourik}\ \emph {et~al.}(2012)\citenamefont {Mourik}, \citenamefont {Zuo}, \citenamefont {Frolov}, \citenamefont {Plissard}, \citenamefont {Bakkers},\ and\ \citenamefont {Kouwenhoven}}]{mourik2012signatures}%
  \BibitemOpen
  \bibfield  {author} {\bibinfo {author} {\bibfnamefont {V.}~\bibnamefont {Mourik}}, \bibinfo {author} {\bibfnamefont {K.}~\bibnamefont {Zuo}}, \bibinfo {author} {\bibfnamefont {S.~M.}\ \bibnamefont {Frolov}}, \bibinfo {author} {\bibfnamefont {S.}~\bibnamefont {Plissard}}, \bibinfo {author} {\bibfnamefont {E.~P.}\ \bibnamefont {Bakkers}},\ and\ \bibinfo {author} {\bibfnamefont {L.~P.}\ \bibnamefont {Kouwenhoven}},\ }\bibfield  {title} {\bibinfo {title} {Signatures of majorana fermions in hybrid superconductor-semiconductor nanowire devices},\ }\href@noop {} {\bibfield  {journal} {\bibinfo  {journal} {Science}\ }\textbf {\bibinfo {volume} {336}},\ \bibinfo {pages} {1003} (\bibinfo {year} {2012})}\BibitemShut {NoStop}%
\bibitem [{\citenamefont {Deng}\ \emph {et~al.}(2012)\citenamefont {Deng}, \citenamefont {Yu}, \citenamefont {Huang}, \citenamefont {Larsson}, \citenamefont {Caroff},\ and\ \citenamefont {Xu}}]{Deng2012}%
  \BibitemOpen
  \bibfield  {author} {\bibinfo {author} {\bibfnamefont {M.~T.}\ \bibnamefont {Deng}}, \bibinfo {author} {\bibfnamefont {C.~L.}\ \bibnamefont {Yu}}, \bibinfo {author} {\bibfnamefont {G.~Y.}\ \bibnamefont {Huang}}, \bibinfo {author} {\bibfnamefont {M.}~\bibnamefont {Larsson}}, \bibinfo {author} {\bibfnamefont {P.}~\bibnamefont {Caroff}},\ and\ \bibinfo {author} {\bibfnamefont {H.~Q.}\ \bibnamefont {Xu}},\ }\bibfield  {title} {\bibinfo {title} {Anomalous zero-bias conductance peak in a nb--insb nanowire--nb hybrid device},\ }\href {https://doi.org/10.1021/nl303758w} {\bibfield  {journal} {\bibinfo  {journal} {Nano Letters}\ }\textbf {\bibinfo {volume} {12}},\ \bibinfo {pages} {6414} (\bibinfo {year} {2012})}\BibitemShut {NoStop}%
\bibitem [{\citenamefont {Das}\ \emph {et~al.}(2012)\citenamefont {Das}, \citenamefont {Ronen}, \citenamefont {Most}, \citenamefont {Oreg}, \citenamefont {Heiblum},\ and\ \citenamefont {Shtrikman}}]{Das2012}%
  \BibitemOpen
  \bibfield  {author} {\bibinfo {author} {\bibfnamefont {A.}~\bibnamefont {Das}}, \bibinfo {author} {\bibfnamefont {Y.}~\bibnamefont {Ronen}}, \bibinfo {author} {\bibfnamefont {Y.}~\bibnamefont {Most}}, \bibinfo {author} {\bibfnamefont {Y.}~\bibnamefont {Oreg}}, \bibinfo {author} {\bibfnamefont {M.}~\bibnamefont {Heiblum}},\ and\ \bibinfo {author} {\bibfnamefont {H.}~\bibnamefont {Shtrikman}},\ }\bibfield  {title} {\bibinfo {title} {Zero-bias peaks and splitting in an al--inas nanowire topological superconductor as a signature of majorana fermions},\ }\href {https://doi.org/10.1038/nphys2479} {\bibfield  {journal} {\bibinfo  {journal} {Nature Physics}\ }\textbf {\bibinfo {volume} {8}},\ \bibinfo {pages} {887} (\bibinfo {year} {2012})}\BibitemShut {NoStop}%
\bibitem [{\citenamefont {Rokhinson}\ \emph {et~al.}(2012)\citenamefont {Rokhinson}, \citenamefont {Liu},\ and\ \citenamefont {Furdyna}}]{rokhinson2012fractional}%
  \BibitemOpen
  \bibfield  {author} {\bibinfo {author} {\bibfnamefont {L.~P.}\ \bibnamefont {Rokhinson}}, \bibinfo {author} {\bibfnamefont {X.}~\bibnamefont {Liu}},\ and\ \bibinfo {author} {\bibfnamefont {J.~K.}\ \bibnamefont {Furdyna}},\ }\bibfield  {title} {\bibinfo {title} {The fractional ac josephson effect in a semiconductor--superconductor nanowire as a signature of majorana particles},\ }\href@noop {} {\bibfield  {journal} {\bibinfo  {journal} {Nature Physics}\ }\textbf {\bibinfo {volume} {8}},\ \bibinfo {pages} {795} (\bibinfo {year} {2012})}\BibitemShut {NoStop}%
\bibitem [{\citenamefont {Churchill}\ \emph {et~al.}(2013)\citenamefont {Churchill}, \citenamefont {Fatemi}, \citenamefont {Grove-Rasmussen}, \citenamefont {Deng}, \citenamefont {Caroff}, \citenamefont {Xu},\ and\ \citenamefont {Marcus}}]{churchill2013superconductor}%
  \BibitemOpen
  \bibfield  {author} {\bibinfo {author} {\bibfnamefont {H.}~\bibnamefont {Churchill}}, \bibinfo {author} {\bibfnamefont {V.}~\bibnamefont {Fatemi}}, \bibinfo {author} {\bibfnamefont {K.}~\bibnamefont {Grove-Rasmussen}}, \bibinfo {author} {\bibfnamefont {M.}~\bibnamefont {Deng}}, \bibinfo {author} {\bibfnamefont {P.}~\bibnamefont {Caroff}}, \bibinfo {author} {\bibfnamefont {H.}~\bibnamefont {Xu}},\ and\ \bibinfo {author} {\bibfnamefont {C.~M.}\ \bibnamefont {Marcus}},\ }\bibfield  {title} {\bibinfo {title} {Superconductor-nanowire devices from tunneling to the multichannel regime: Zero-bias oscillations and magnetoconductance crossover},\ }\href@noop {} {\bibfield  {journal} {\bibinfo  {journal} {Physical Review B}\ }\textbf {\bibinfo {volume} {87}},\ \bibinfo {pages} {241401} (\bibinfo {year} {2013})}\BibitemShut {NoStop}%
\bibitem [{\citenamefont {Finck}\ \emph {et~al.}(2013)\citenamefont {Finck}, \citenamefont {Van~Harlingen}, \citenamefont {Mohseni}, \citenamefont {Jung},\ and\ \citenamefont {Li}}]{finck2013anomalous}%
  \BibitemOpen
  \bibfield  {author} {\bibinfo {author} {\bibfnamefont {A.}~\bibnamefont {Finck}}, \bibinfo {author} {\bibfnamefont {D.~J.}\ \bibnamefont {Van~Harlingen}}, \bibinfo {author} {\bibfnamefont {P.}~\bibnamefont {Mohseni}}, \bibinfo {author} {\bibfnamefont {K.}~\bibnamefont {Jung}},\ and\ \bibinfo {author} {\bibfnamefont {X.}~\bibnamefont {Li}},\ }\bibfield  {title} {\bibinfo {title} {Anomalous modulation of a zero-bias peak in a hybrid nanowire-superconductor device},\ }\href@noop {} {\bibfield  {journal} {\bibinfo  {journal} {Physical review letters}\ }\textbf {\bibinfo {volume} {110}},\ \bibinfo {pages} {126406} (\bibinfo {year} {2013})}\BibitemShut {NoStop}%
\bibitem [{\citenamefont {Deng}\ \emph {et~al.}(2016)\citenamefont {Deng}, \citenamefont {Vaitiek{\.e}nas}, \citenamefont {Hansen}, \citenamefont {Danon}, \citenamefont {Leijnse}, \citenamefont {Flensberg}, \citenamefont {Nyg{\aa}rd}, \citenamefont {Krogstrup},\ and\ \citenamefont {Marcus}}]{deng2016majorana}%
  \BibitemOpen
  \bibfield  {author} {\bibinfo {author} {\bibfnamefont {M.}~\bibnamefont {Deng}}, \bibinfo {author} {\bibfnamefont {S.}~\bibnamefont {Vaitiek{\.e}nas}}, \bibinfo {author} {\bibfnamefont {E.~B.}\ \bibnamefont {Hansen}}, \bibinfo {author} {\bibfnamefont {J.}~\bibnamefont {Danon}}, \bibinfo {author} {\bibfnamefont {M.}~\bibnamefont {Leijnse}}, \bibinfo {author} {\bibfnamefont {K.}~\bibnamefont {Flensberg}}, \bibinfo {author} {\bibfnamefont {J.}~\bibnamefont {Nyg{\aa}rd}}, \bibinfo {author} {\bibfnamefont {P.}~\bibnamefont {Krogstrup}},\ and\ \bibinfo {author} {\bibfnamefont {C.~M.}\ \bibnamefont {Marcus}},\ }\bibfield  {title} {\bibinfo {title} {Majorana bound state in a coupled quantum-dot hybrid-nanowire system},\ }\href@noop {} {\bibfield  {journal} {\bibinfo  {journal} {Science}\ }\textbf {\bibinfo {volume} {354}},\ \bibinfo {pages} {1557} (\bibinfo {year} {2016})}\BibitemShut {NoStop}%
\bibitem [{\citenamefont {Zhang}\ \emph {et~al.}(2017)\citenamefont {Zhang}, \citenamefont {G{\"u}l}, \citenamefont {Conesa-Boj}, \citenamefont {Nowak}, \citenamefont {Wimmer}, \citenamefont {Zuo}, \citenamefont {Mourik}, \citenamefont {De~Vries}, \citenamefont {Van~Veen}, \citenamefont {De~Moor} \emph {et~al.}}]{zhang2017ballistic}%
  \BibitemOpen
  \bibfield  {author} {\bibinfo {author} {\bibfnamefont {H.}~\bibnamefont {Zhang}}, \bibinfo {author} {\bibfnamefont {{\"O}.}~\bibnamefont {G{\"u}l}}, \bibinfo {author} {\bibfnamefont {S.}~\bibnamefont {Conesa-Boj}}, \bibinfo {author} {\bibfnamefont {M.~P.}\ \bibnamefont {Nowak}}, \bibinfo {author} {\bibfnamefont {M.}~\bibnamefont {Wimmer}}, \bibinfo {author} {\bibfnamefont {K.}~\bibnamefont {Zuo}}, \bibinfo {author} {\bibfnamefont {V.}~\bibnamefont {Mourik}}, \bibinfo {author} {\bibfnamefont {F.~K.}\ \bibnamefont {De~Vries}}, \bibinfo {author} {\bibfnamefont {J.}~\bibnamefont {Van~Veen}}, \bibinfo {author} {\bibfnamefont {M.~W.}\ \bibnamefont {De~Moor}}, \emph {et~al.},\ }\bibfield  {title} {\bibinfo {title} {Ballistic superconductivity in semiconductor nanowires},\ }\href@noop {} {\bibfield  {journal} {\bibinfo  {journal} {Nature communications}\ }\textbf {\bibinfo {volume} {8}},\ \bibinfo {pages} {16025} (\bibinfo {year} {2017})}\BibitemShut {NoStop}%
\bibitem [{\citenamefont {Chen}\ \emph {et~al.}(2017)\citenamefont {Chen}, \citenamefont {Yu}, \citenamefont {Stenger}, \citenamefont {Hocevar}, \citenamefont {Car}, \citenamefont {Plissard}, \citenamefont {Bakkers}, \citenamefont {Stanescu},\ and\ \citenamefont {Frolov}}]{chen2017experimental}%
  \BibitemOpen
  \bibfield  {author} {\bibinfo {author} {\bibfnamefont {J.}~\bibnamefont {Chen}}, \bibinfo {author} {\bibfnamefont {P.}~\bibnamefont {Yu}}, \bibinfo {author} {\bibfnamefont {J.}~\bibnamefont {Stenger}}, \bibinfo {author} {\bibfnamefont {M.}~\bibnamefont {Hocevar}}, \bibinfo {author} {\bibfnamefont {D.}~\bibnamefont {Car}}, \bibinfo {author} {\bibfnamefont {S.~R.}\ \bibnamefont {Plissard}}, \bibinfo {author} {\bibfnamefont {E.~P.}\ \bibnamefont {Bakkers}}, \bibinfo {author} {\bibfnamefont {T.~D.}\ \bibnamefont {Stanescu}},\ and\ \bibinfo {author} {\bibfnamefont {S.~M.}\ \bibnamefont {Frolov}},\ }\bibfield  {title} {\bibinfo {title} {Experimental phase diagram of zero-bias conductance peaks in superconductor/semiconductor nanowire devices},\ }\href@noop {} {\bibfield  {journal} {\bibinfo  {journal} {Science advances}\ }\textbf {\bibinfo {volume} {3}},\ \bibinfo {pages} {e1701476} (\bibinfo {year} {2017})}\BibitemShut {NoStop}%
\bibitem [{\citenamefont {Nichele}\ \emph {et~al.}(2017)\citenamefont {Nichele}, \citenamefont {Drachmann}, \citenamefont {Whiticar}, \citenamefont {O’Farrell}, \citenamefont {Suominen}, \citenamefont {Fornieri}, \citenamefont {Wang}, \citenamefont {Gardner}, \citenamefont {Thomas}, \citenamefont {Hatke} \emph {et~al.}}]{nichele2017scaling}%
  \BibitemOpen
  \bibfield  {author} {\bibinfo {author} {\bibfnamefont {F.}~\bibnamefont {Nichele}}, \bibinfo {author} {\bibfnamefont {A.~C.}\ \bibnamefont {Drachmann}}, \bibinfo {author} {\bibfnamefont {A.~M.}\ \bibnamefont {Whiticar}}, \bibinfo {author} {\bibfnamefont {E.~C.}\ \bibnamefont {O’Farrell}}, \bibinfo {author} {\bibfnamefont {H.~J.}\ \bibnamefont {Suominen}}, \bibinfo {author} {\bibfnamefont {A.}~\bibnamefont {Fornieri}}, \bibinfo {author} {\bibfnamefont {T.}~\bibnamefont {Wang}}, \bibinfo {author} {\bibfnamefont {G.~C.}\ \bibnamefont {Gardner}}, \bibinfo {author} {\bibfnamefont {C.}~\bibnamefont {Thomas}}, \bibinfo {author} {\bibfnamefont {A.~T.}\ \bibnamefont {Hatke}}, \emph {et~al.},\ }\bibfield  {title} {\bibinfo {title} {Scaling of majorana zero-bias conductance peaks},\ }\href@noop {} {\bibfield  {journal} {\bibinfo  {journal} {Physical review letters}\ }\textbf {\bibinfo {volume} {119}},\ \bibinfo {pages} {136803} (\bibinfo {year} {2017})}\BibitemShut {NoStop}%
\bibitem [{\citenamefont {Albrecht}\ \emph {et~al.}(2017)\citenamefont {Albrecht}, \citenamefont {Hansen}, \citenamefont {Higginbotham}, \citenamefont {Kuemmeth}, \citenamefont {Jespersen}, \citenamefont {Nyg{\aa}rd}, \citenamefont {Krogstrup}, \citenamefont {Danon}, \citenamefont {Flensberg},\ and\ \citenamefont {Marcus}}]{albrecht2017transport}%
  \BibitemOpen
  \bibfield  {author} {\bibinfo {author} {\bibfnamefont {S.}~\bibnamefont {Albrecht}}, \bibinfo {author} {\bibfnamefont {E.}~\bibnamefont {Hansen}}, \bibinfo {author} {\bibfnamefont {A.~P.}\ \bibnamefont {Higginbotham}}, \bibinfo {author} {\bibfnamefont {F.}~\bibnamefont {Kuemmeth}}, \bibinfo {author} {\bibfnamefont {T.}~\bibnamefont {Jespersen}}, \bibinfo {author} {\bibfnamefont {J.}~\bibnamefont {Nyg{\aa}rd}}, \bibinfo {author} {\bibfnamefont {P.}~\bibnamefont {Krogstrup}}, \bibinfo {author} {\bibfnamefont {J.}~\bibnamefont {Danon}}, \bibinfo {author} {\bibfnamefont {K.}~\bibnamefont {Flensberg}},\ and\ \bibinfo {author} {\bibfnamefont {C.}~\bibnamefont {Marcus}},\ }\bibfield  {title} {\bibinfo {title} {Transport signatures of quasiparticle poisoning in a majorana island},\ }\href@noop {} {\bibfield  {journal} {\bibinfo  {journal} {Physical review letters}\ }\textbf {\bibinfo {volume} {118}},\ \bibinfo {pages} {137701} (\bibinfo {year} {2017})}\BibitemShut {NoStop}%
\bibitem [{\citenamefont {O~Farrell}\ \emph {et~al.}(2018)\citenamefont {O~Farrell}, \citenamefont {Drachmann}, \citenamefont {Hell}, \citenamefont {Fornieri}, \citenamefont {Whiticar}, \citenamefont {Hansen}, \citenamefont {Gronin}, \citenamefont {Gardner}, \citenamefont {Thomas}, \citenamefont {Manfra} \emph {et~al.}}]{o2018hybridization}%
  \BibitemOpen
  \bibfield  {author} {\bibinfo {author} {\bibfnamefont {E.}~\bibnamefont {O~Farrell}}, \bibinfo {author} {\bibfnamefont {A.}~\bibnamefont {Drachmann}}, \bibinfo {author} {\bibfnamefont {M.}~\bibnamefont {Hell}}, \bibinfo {author} {\bibfnamefont {A.}~\bibnamefont {Fornieri}}, \bibinfo {author} {\bibfnamefont {A.}~\bibnamefont {Whiticar}}, \bibinfo {author} {\bibfnamefont {E.}~\bibnamefont {Hansen}}, \bibinfo {author} {\bibfnamefont {S.}~\bibnamefont {Gronin}}, \bibinfo {author} {\bibfnamefont {G.}~\bibnamefont {Gardner}}, \bibinfo {author} {\bibfnamefont {C.}~\bibnamefont {Thomas}}, \bibinfo {author} {\bibfnamefont {M.}~\bibnamefont {Manfra}}, \emph {et~al.},\ }\bibfield  {title} {\bibinfo {title} {Hybridization of subgap states in one-dimensional superconductor-semiconductor coulomb islands},\ }\href@noop {} {\bibfield  {journal} {\bibinfo  {journal} {Physical review letters}\ }\textbf {\bibinfo {volume} {121}},\ \bibinfo {pages} {256803} (\bibinfo {year} {2018})}\BibitemShut {NoStop}%
\bibitem [{\citenamefont {Shen}\ \emph {et~al.}(2018)\citenamefont {Shen}, \citenamefont {Heedt}, \citenamefont {Borsoi}, \citenamefont {Van~Heck}, \citenamefont {Gazibegovic}, \citenamefont {het Veld}, \citenamefont {Car}, \citenamefont {Logan}, \citenamefont {Pendharkar}, \citenamefont {Ramakers} \emph {et~al.}}]{shen2018parity}%
  \BibitemOpen
  \bibfield  {author} {\bibinfo {author} {\bibfnamefont {J.}~\bibnamefont {Shen}}, \bibinfo {author} {\bibfnamefont {S.}~\bibnamefont {Heedt}}, \bibinfo {author} {\bibfnamefont {F.}~\bibnamefont {Borsoi}}, \bibinfo {author} {\bibfnamefont {B.}~\bibnamefont {Van~Heck}}, \bibinfo {author} {\bibfnamefont {S.}~\bibnamefont {Gazibegovic}}, \bibinfo {author} {\bibfnamefont {R.~L.~O.}\ \bibnamefont {het Veld}}, \bibinfo {author} {\bibfnamefont {D.}~\bibnamefont {Car}}, \bibinfo {author} {\bibfnamefont {J.~A.}\ \bibnamefont {Logan}}, \bibinfo {author} {\bibfnamefont {M.}~\bibnamefont {Pendharkar}}, \bibinfo {author} {\bibfnamefont {S.~J.}\ \bibnamefont {Ramakers}}, \emph {et~al.},\ }\bibfield  {title} {\bibinfo {title} {Parity transitions in the superconducting ground state of hybrid insb--al coulomb islands},\ }\href@noop {} {\bibfield  {journal} {\bibinfo  {journal} {Nature communications}\ }\textbf {\bibinfo {volume} {9}},\ \bibinfo {pages} {4801} (\bibinfo {year} {2018})}\BibitemShut {NoStop}%
\bibitem [{\citenamefont {Sherman}\ \emph {et~al.}(2017)\citenamefont {Sherman}, \citenamefont {Yodh}, \citenamefont {Albrecht}, \citenamefont {Nyg{\aa}rd}, \citenamefont {Krogstrup},\ and\ \citenamefont {Marcus}}]{sherman2017normal}%
  \BibitemOpen
  \bibfield  {author} {\bibinfo {author} {\bibfnamefont {D.}~\bibnamefont {Sherman}}, \bibinfo {author} {\bibfnamefont {J.}~\bibnamefont {Yodh}}, \bibinfo {author} {\bibfnamefont {S.~M.}\ \bibnamefont {Albrecht}}, \bibinfo {author} {\bibfnamefont {J.}~\bibnamefont {Nyg{\aa}rd}}, \bibinfo {author} {\bibfnamefont {P.}~\bibnamefont {Krogstrup}},\ and\ \bibinfo {author} {\bibfnamefont {C.~M.}\ \bibnamefont {Marcus}},\ }\bibfield  {title} {\bibinfo {title} {Normal, superconducting and topological regimes of hybrid double quantum dots},\ }\href@noop {} {\bibfield  {journal} {\bibinfo  {journal} {Nature nanotechnology}\ }\textbf {\bibinfo {volume} {12}},\ \bibinfo {pages} {212} (\bibinfo {year} {2017})}\BibitemShut {NoStop}%
\bibitem [{\citenamefont {Vaitiek{\.e}nas}\ \emph {et~al.}(2018)\citenamefont {Vaitiek{\.e}nas}, \citenamefont {Whiticar}, \citenamefont {Deng}, \citenamefont {Krizek}, \citenamefont {Sestoft}, \citenamefont {Palmstr{\o}m}, \citenamefont {Marti-Sanchez}, \citenamefont {Arbiol}, \citenamefont {Krogstrup}, \citenamefont {Casparis} \emph {et~al.}}]{vaitiekenas2018selective}%
  \BibitemOpen
  \bibfield  {author} {\bibinfo {author} {\bibfnamefont {S.}~\bibnamefont {Vaitiek{\.e}nas}}, \bibinfo {author} {\bibfnamefont {A.}~\bibnamefont {Whiticar}}, \bibinfo {author} {\bibfnamefont {M.-T.}\ \bibnamefont {Deng}}, \bibinfo {author} {\bibfnamefont {F.}~\bibnamefont {Krizek}}, \bibinfo {author} {\bibfnamefont {J.}~\bibnamefont {Sestoft}}, \bibinfo {author} {\bibfnamefont {C.}~\bibnamefont {Palmstr{\o}m}}, \bibinfo {author} {\bibfnamefont {S.}~\bibnamefont {Marti-Sanchez}}, \bibinfo {author} {\bibfnamefont {J.}~\bibnamefont {Arbiol}}, \bibinfo {author} {\bibfnamefont {P.}~\bibnamefont {Krogstrup}}, \bibinfo {author} {\bibfnamefont {L.}~\bibnamefont {Casparis}}, \emph {et~al.},\ }\bibfield  {title} {\bibinfo {title} {Selective-area-grown semiconductor-superconductor hybrids: A basis for topological networks},\ }\href@noop {} {\bibfield  {journal} {\bibinfo  {journal} {Physical review letters}\ }\textbf {\bibinfo {volume} {121}},\ \bibinfo {pages} {147701} (\bibinfo {year} {2018})}\BibitemShut {NoStop}%
\bibitem [{\citenamefont {Albrecht}\ \emph {et~al.}(2016)\citenamefont {Albrecht}, \citenamefont {Higginbotham}, \citenamefont {Madsen}, \citenamefont {Kuemmeth}, \citenamefont {Jespersen}, \citenamefont {Nyg{\aa}rd}, \citenamefont {Krogstrup},\ and\ \citenamefont {Marcus}}]{albrecht2016exponential}%
  \BibitemOpen
  \bibfield  {author} {\bibinfo {author} {\bibfnamefont {S.~M.}\ \bibnamefont {Albrecht}}, \bibinfo {author} {\bibfnamefont {A.~P.}\ \bibnamefont {Higginbotham}}, \bibinfo {author} {\bibfnamefont {M.}~\bibnamefont {Madsen}}, \bibinfo {author} {\bibfnamefont {F.}~\bibnamefont {Kuemmeth}}, \bibinfo {author} {\bibfnamefont {T.~S.}\ \bibnamefont {Jespersen}}, \bibinfo {author} {\bibfnamefont {J.}~\bibnamefont {Nyg{\aa}rd}}, \bibinfo {author} {\bibfnamefont {P.}~\bibnamefont {Krogstrup}},\ and\ \bibinfo {author} {\bibfnamefont {C.}~\bibnamefont {Marcus}},\ }\bibfield  {title} {\bibinfo {title} {Exponential protection of zero modes in majorana islands},\ }\href@noop {} {\bibfield  {journal} {\bibinfo  {journal} {Nature}\ }\textbf {\bibinfo {volume} {531}},\ \bibinfo {pages} {206} (\bibinfo {year} {2016})}\BibitemShut {NoStop}%
\bibitem [{\citenamefont {Yu}\ \emph {et~al.}(2021)\citenamefont {Yu}, \citenamefont {Chen}, \citenamefont {Gomanko}, \citenamefont {Badawy}, \citenamefont {Bakkers}, \citenamefont {Zuo}, \citenamefont {Mourik},\ and\ \citenamefont {Frolov}}]{Yu_2021}%
  \BibitemOpen
  \bibfield  {author} {\bibinfo {author} {\bibfnamefont {P.}~\bibnamefont {Yu}}, \bibinfo {author} {\bibfnamefont {J.}~\bibnamefont {Chen}}, \bibinfo {author} {\bibfnamefont {M.}~\bibnamefont {Gomanko}}, \bibinfo {author} {\bibfnamefont {G.}~\bibnamefont {Badawy}}, \bibinfo {author} {\bibfnamefont {E.~P. A.~M.}\ \bibnamefont {Bakkers}}, \bibinfo {author} {\bibfnamefont {K.}~\bibnamefont {Zuo}}, \bibinfo {author} {\bibfnamefont {V.}~\bibnamefont {Mourik}},\ and\ \bibinfo {author} {\bibfnamefont {S.~M.}\ \bibnamefont {Frolov}},\ }\bibfield  {title} {\bibinfo {title} {Non-majorana states yield nearly quantized conductance in proximatized nanowires},\ }\href {https://doi.org/10.1038/s41567-020-01107-w} {\bibfield  {journal} {\bibinfo  {journal} {Nature Physics}\ }\textbf {\bibinfo {volume} {17}},\ \bibinfo {pages} {482–488} (\bibinfo {year} {2021})}\BibitemShut {NoStop}%
\bibitem [{\citenamefont {Zhang}\ \emph {et~al.}(2021)\citenamefont {Zhang}, \citenamefont {de~Moor}, \citenamefont {Bommer}, \citenamefont {Xu}, \citenamefont {Wang}, \citenamefont {van Loo}, \citenamefont {Liu}, \citenamefont {Gazibegovic}, \citenamefont {Logan}, \citenamefont {Car}, \citenamefont {het Veld}, \citenamefont {van Veldhoven}, \citenamefont {Koelling}, \citenamefont {Verheijen}, \citenamefont {Pendharkar}, \citenamefont {Pennachio}, \citenamefont {Shojaei}, \citenamefont {Lee}, \citenamefont {Palmstrøm}, \citenamefont {Bakkers}, \citenamefont {Sarma},\ and\ \citenamefont {Kouwenhoven}}]{zhang2021}%
  \BibitemOpen
  \bibfield  {author} {\bibinfo {author} {\bibfnamefont {H.}~\bibnamefont {Zhang}}, \bibinfo {author} {\bibfnamefont {M.~W.~A.}\ \bibnamefont {de~Moor}}, \bibinfo {author} {\bibfnamefont {J.~D.~S.}\ \bibnamefont {Bommer}}, \bibinfo {author} {\bibfnamefont {D.}~\bibnamefont {Xu}}, \bibinfo {author} {\bibfnamefont {G.}~\bibnamefont {Wang}}, \bibinfo {author} {\bibfnamefont {N.}~\bibnamefont {van Loo}}, \bibinfo {author} {\bibfnamefont {C.-X.}\ \bibnamefont {Liu}}, \bibinfo {author} {\bibfnamefont {S.}~\bibnamefont {Gazibegovic}}, \bibinfo {author} {\bibfnamefont {J.~A.}\ \bibnamefont {Logan}}, \bibinfo {author} {\bibfnamefont {D.}~\bibnamefont {Car}}, \bibinfo {author} {\bibfnamefont {R.~L. M.~O.}\ \bibnamefont {het Veld}}, \bibinfo {author} {\bibfnamefont {P.~J.}\ \bibnamefont {van Veldhoven}}, \bibinfo {author} {\bibfnamefont {S.}~\bibnamefont {Koelling}}, \bibinfo {author} {\bibfnamefont {M.~A.}\ \bibnamefont {Verheijen}}, \bibinfo {author} {\bibfnamefont {M.}~\bibnamefont {Pendharkar}}, \bibinfo {author}
  {\bibfnamefont {D.~J.}\ \bibnamefont {Pennachio}}, \bibinfo {author} {\bibfnamefont {B.}~\bibnamefont {Shojaei}}, \bibinfo {author} {\bibfnamefont {J.~S.}\ \bibnamefont {Lee}}, \bibinfo {author} {\bibfnamefont {C.~J.}\ \bibnamefont {Palmstrøm}}, \bibinfo {author} {\bibfnamefont {E.~P. A.~M.}\ \bibnamefont {Bakkers}}, \bibinfo {author} {\bibfnamefont {S.~D.}\ \bibnamefont {Sarma}},\ and\ \bibinfo {author} {\bibfnamefont {L.~P.}\ \bibnamefont {Kouwenhoven}},\ }\href@noop {} {\bibinfo {title} {Large zero-bias peaks in insb-al hybrid semiconductor-superconductor nanowire devices}} (\bibinfo {year} {2021}),\ \Eprint {https://arxiv.org/abs/2101.11456} {arXiv:2101.11456 [cond-mat.mes-hall]} \BibitemShut {NoStop}%
\bibitem [{\citenamefont {Aghaee}\ \emph {et~al.}(2023)\citenamefont {Aghaee}, \citenamefont {Akkala}, \citenamefont {Alam}, \citenamefont {Ali}, \citenamefont {Alcaraz~Ramirez}, \citenamefont {Andrzejczuk}, \citenamefont {Antipov}, \citenamefont {Aseev}, \citenamefont {Astafev}, \citenamefont {Bauer}, \citenamefont {Becker}, \citenamefont {Boddapati}, \citenamefont {Boekhout}, \citenamefont {Bommer}, \citenamefont {Bosma}, \citenamefont {Bourdet}, \citenamefont {Boutin}, \citenamefont {Caroff}, \citenamefont {Casparis}, \citenamefont {Cassidy}, \citenamefont {Chatoor}, \citenamefont {Christensen}, \citenamefont {Clay}, \citenamefont {Cole}, \citenamefont {Corsetti}, \citenamefont {Cui}, \citenamefont {Dalampiras}, \citenamefont {Dokania}, \citenamefont {de~Lange}, \citenamefont {de~Moor}, \citenamefont {Estrada Salda\~na}, \citenamefont {Fallahi}, \citenamefont {Fathabad}, \citenamefont {Gamble}, \citenamefont {Gardner}, \citenamefont {Govender}, \citenamefont {Griggio}, \citenamefont {Grigoryan},
  \citenamefont {Gronin}, \citenamefont {Gukelberger}, \citenamefont {Hansen}, \citenamefont {Heedt}, \citenamefont {Herranz~Zamorano}, \citenamefont {Ho}, \citenamefont {Holgaard}, \citenamefont {Ingerslev}, \citenamefont {Johansson}, \citenamefont {Jones}, \citenamefont {Kallaher}, \citenamefont {Karimi}, \citenamefont {Karzig}, \citenamefont {King}, \citenamefont {Kloster}, \citenamefont {Knapp}, \citenamefont {Kocon}, \citenamefont {Koski}, \citenamefont {Kostamo}, \citenamefont {Krogstrup}, \citenamefont {Kumar}, \citenamefont {Laeven}, \citenamefont {Larsen}, \citenamefont {Li}, \citenamefont {Lindemann}, \citenamefont {Love}, \citenamefont {Lutchyn}, \citenamefont {Madsen}, \citenamefont {Manfra}, \citenamefont {Markussen}, \citenamefont {Martinez}, \citenamefont {McNeil}, \citenamefont {Memisevic}, \citenamefont {Morgan}, \citenamefont {Mullally}, \citenamefont {Nayak}, \citenamefont {Nielsen}, \citenamefont {Nielsen}, \citenamefont {Nijholt}, \citenamefont {Nurmohamed}, \citenamefont {O'Farrell},
  \citenamefont {Otani}, \citenamefont {Pauka}, \citenamefont {Petersson}, \citenamefont {Petit}, \citenamefont {Pikulin}, \citenamefont {Preiss}, \citenamefont {Quintero-Perez}, \citenamefont {Rajpalke}, \citenamefont {Rasmussen}, \citenamefont {Razmadze}, \citenamefont {Reentila}, \citenamefont {Reilly}, \citenamefont {Rouse}, \citenamefont {Sadovskyy}, \citenamefont {Sainiemi}, \citenamefont {Schreppler}, \citenamefont {Sidorkin}, \citenamefont {Singh}, \citenamefont {Singh}, \citenamefont {Sinha}, \citenamefont {Sohr}, \citenamefont {Stankevi\ifmmode~\check{c}\else \v{c}\fi{}}, \citenamefont {Stek}, \citenamefont {Suominen}, \citenamefont {Suter}, \citenamefont {Svidenko}, \citenamefont {Teicher}, \citenamefont {Temuerhan}, \citenamefont {Thiyagarajah}, \citenamefont {Tholapi}, \citenamefont {Thomas}, \citenamefont {Toomey}, \citenamefont {Upadhyay}, \citenamefont {Urban}, \citenamefont {Vaitiek\ifmmode~\dot{e}\else \.{e}\fi{}nas}, \citenamefont {Van~Hoogdalem}, \citenamefont {Van~Woerkom}, \citenamefont
  {Viazmitinov}, \citenamefont {Vogel}, \citenamefont {Waddy}, \citenamefont {Watson}, \citenamefont {Weston}, \citenamefont {Winkler}, \citenamefont {Yang}, \citenamefont {Yau}, \citenamefont {Yi}, \citenamefont {Yucelen}, \citenamefont {Webster}, \citenamefont {Zeisel},\ and\ \citenamefont {Zhao}}]{PhysRevB.107.245423}%
  \BibitemOpen
  \bibfield  {author} {\bibinfo {author} {\bibfnamefont {M.}~\bibnamefont {Aghaee}}, \bibinfo {author} {\bibfnamefont {A.}~\bibnamefont {Akkala}}, \bibinfo {author} {\bibfnamefont {Z.}~\bibnamefont {Alam}}, \bibinfo {author} {\bibfnamefont {R.}~\bibnamefont {Ali}}, \bibinfo {author} {\bibfnamefont {A.}~\bibnamefont {Alcaraz~Ramirez}}, \bibinfo {author} {\bibfnamefont {M.}~\bibnamefont {Andrzejczuk}}, \bibinfo {author} {\bibfnamefont {A.~E.}\ \bibnamefont {Antipov}}, \bibinfo {author} {\bibfnamefont {P.}~\bibnamefont {Aseev}}, \bibinfo {author} {\bibfnamefont {M.}~\bibnamefont {Astafev}}, \bibinfo {author} {\bibfnamefont {B.}~\bibnamefont {Bauer}}, \bibinfo {author} {\bibfnamefont {J.}~\bibnamefont {Becker}}, \bibinfo {author} {\bibfnamefont {S.}~\bibnamefont {Boddapati}}, \bibinfo {author} {\bibfnamefont {F.}~\bibnamefont {Boekhout}}, \bibinfo {author} {\bibfnamefont {J.}~\bibnamefont {Bommer}}, \bibinfo {author} {\bibfnamefont {T.}~\bibnamefont {Bosma}}, \bibinfo {author} {\bibfnamefont {L.}~\bibnamefont
  {Bourdet}}, \bibinfo {author} {\bibfnamefont {S.}~\bibnamefont {Boutin}}, \bibinfo {author} {\bibfnamefont {P.}~\bibnamefont {Caroff}}, \bibinfo {author} {\bibfnamefont {L.}~\bibnamefont {Casparis}}, \bibinfo {author} {\bibfnamefont {M.}~\bibnamefont {Cassidy}}, \bibinfo {author} {\bibfnamefont {S.}~\bibnamefont {Chatoor}}, \bibinfo {author} {\bibfnamefont {A.~W.}\ \bibnamefont {Christensen}}, \bibinfo {author} {\bibfnamefont {N.}~\bibnamefont {Clay}}, \bibinfo {author} {\bibfnamefont {W.~S.}\ \bibnamefont {Cole}}, \bibinfo {author} {\bibfnamefont {F.}~\bibnamefont {Corsetti}}, \bibinfo {author} {\bibfnamefont {A.}~\bibnamefont {Cui}}, \bibinfo {author} {\bibfnamefont {P.}~\bibnamefont {Dalampiras}}, \bibinfo {author} {\bibfnamefont {A.}~\bibnamefont {Dokania}}, \bibinfo {author} {\bibfnamefont {G.}~\bibnamefont {de~Lange}}, \bibinfo {author} {\bibfnamefont {M.}~\bibnamefont {de~Moor}}, \bibinfo {author} {\bibfnamefont {J.~C.}\ \bibnamefont {Estrada Salda\~na}}, \bibinfo {author} {\bibfnamefont
  {S.}~\bibnamefont {Fallahi}}, \bibinfo {author} {\bibfnamefont {Z.~H.}\ \bibnamefont {Fathabad}}, \bibinfo {author} {\bibfnamefont {J.}~\bibnamefont {Gamble}}, \bibinfo {author} {\bibfnamefont {G.}~\bibnamefont {Gardner}}, \bibinfo {author} {\bibfnamefont {D.}~\bibnamefont {Govender}}, \bibinfo {author} {\bibfnamefont {F.}~\bibnamefont {Griggio}}, \bibinfo {author} {\bibfnamefont {R.}~\bibnamefont {Grigoryan}}, \bibinfo {author} {\bibfnamefont {S.}~\bibnamefont {Gronin}}, \bibinfo {author} {\bibfnamefont {J.}~\bibnamefont {Gukelberger}}, \bibinfo {author} {\bibfnamefont {E.~B.}\ \bibnamefont {Hansen}}, \bibinfo {author} {\bibfnamefont {S.}~\bibnamefont {Heedt}}, \bibinfo {author} {\bibfnamefont {J.}~\bibnamefont {Herranz~Zamorano}}, \bibinfo {author} {\bibfnamefont {S.}~\bibnamefont {Ho}}, \bibinfo {author} {\bibfnamefont {U.~L.}\ \bibnamefont {Holgaard}}, \bibinfo {author} {\bibfnamefont {H.}~\bibnamefont {Ingerslev}}, \bibinfo {author} {\bibfnamefont {L.}~\bibnamefont {Johansson}}, \bibinfo {author}
  {\bibfnamefont {J.}~\bibnamefont {Jones}}, \bibinfo {author} {\bibfnamefont {R.}~\bibnamefont {Kallaher}}, \bibinfo {author} {\bibfnamefont {F.}~\bibnamefont {Karimi}}, \bibinfo {author} {\bibfnamefont {T.}~\bibnamefont {Karzig}}, \bibinfo {author} {\bibfnamefont {C.}~\bibnamefont {King}}, \bibinfo {author} {\bibfnamefont {M.~E.}\ \bibnamefont {Kloster}}, \bibinfo {author} {\bibfnamefont {C.}~\bibnamefont {Knapp}}, \bibinfo {author} {\bibfnamefont {D.}~\bibnamefont {Kocon}}, \bibinfo {author} {\bibfnamefont {J.}~\bibnamefont {Koski}}, \bibinfo {author} {\bibfnamefont {P.}~\bibnamefont {Kostamo}}, \bibinfo {author} {\bibfnamefont {P.}~\bibnamefont {Krogstrup}}, \bibinfo {author} {\bibfnamefont {M.}~\bibnamefont {Kumar}}, \bibinfo {author} {\bibfnamefont {T.}~\bibnamefont {Laeven}}, \bibinfo {author} {\bibfnamefont {T.}~\bibnamefont {Larsen}}, \bibinfo {author} {\bibfnamefont {K.}~\bibnamefont {Li}}, \bibinfo {author} {\bibfnamefont {T.}~\bibnamefont {Lindemann}}, \bibinfo {author} {\bibfnamefont
  {J.}~\bibnamefont {Love}}, \bibinfo {author} {\bibfnamefont {R.}~\bibnamefont {Lutchyn}}, \bibinfo {author} {\bibfnamefont {M.~H.}\ \bibnamefont {Madsen}}, \bibinfo {author} {\bibfnamefont {M.}~\bibnamefont {Manfra}}, \bibinfo {author} {\bibfnamefont {S.}~\bibnamefont {Markussen}}, \bibinfo {author} {\bibfnamefont {E.}~\bibnamefont {Martinez}}, \bibinfo {author} {\bibfnamefont {R.}~\bibnamefont {McNeil}}, \bibinfo {author} {\bibfnamefont {E.}~\bibnamefont {Memisevic}}, \bibinfo {author} {\bibfnamefont {T.}~\bibnamefont {Morgan}}, \bibinfo {author} {\bibfnamefont {A.}~\bibnamefont {Mullally}}, \bibinfo {author} {\bibfnamefont {C.}~\bibnamefont {Nayak}}, \bibinfo {author} {\bibfnamefont {J.}~\bibnamefont {Nielsen}}, \bibinfo {author} {\bibfnamefont {W.~H.~P.}\ \bibnamefont {Nielsen}}, \bibinfo {author} {\bibfnamefont {B.}~\bibnamefont {Nijholt}}, \bibinfo {author} {\bibfnamefont {A.}~\bibnamefont {Nurmohamed}}, \bibinfo {author} {\bibfnamefont {E.}~\bibnamefont {O'Farrell}}, \bibinfo {author} {\bibfnamefont
  {K.}~\bibnamefont {Otani}}, \bibinfo {author} {\bibfnamefont {S.}~\bibnamefont {Pauka}}, \bibinfo {author} {\bibfnamefont {K.}~\bibnamefont {Petersson}}, \bibinfo {author} {\bibfnamefont {L.}~\bibnamefont {Petit}}, \bibinfo {author} {\bibfnamefont {D.~I.}\ \bibnamefont {Pikulin}}, \bibinfo {author} {\bibfnamefont {F.}~\bibnamefont {Preiss}}, \bibinfo {author} {\bibfnamefont {M.}~\bibnamefont {Quintero-Perez}}, \bibinfo {author} {\bibfnamefont {M.}~\bibnamefont {Rajpalke}}, \bibinfo {author} {\bibfnamefont {K.}~\bibnamefont {Rasmussen}}, \bibinfo {author} {\bibfnamefont {D.}~\bibnamefont {Razmadze}}, \bibinfo {author} {\bibfnamefont {O.}~\bibnamefont {Reentila}}, \bibinfo {author} {\bibfnamefont {D.}~\bibnamefont {Reilly}}, \bibinfo {author} {\bibfnamefont {R.}~\bibnamefont {Rouse}}, \bibinfo {author} {\bibfnamefont {I.}~\bibnamefont {Sadovskyy}}, \bibinfo {author} {\bibfnamefont {L.}~\bibnamefont {Sainiemi}}, \bibinfo {author} {\bibfnamefont {S.}~\bibnamefont {Schreppler}}, \bibinfo {author} {\bibfnamefont
  {V.}~\bibnamefont {Sidorkin}}, \bibinfo {author} {\bibfnamefont {A.}~\bibnamefont {Singh}}, \bibinfo {author} {\bibfnamefont {S.}~\bibnamefont {Singh}}, \bibinfo {author} {\bibfnamefont {S.}~\bibnamefont {Sinha}}, \bibinfo {author} {\bibfnamefont {P.}~\bibnamefont {Sohr}}, \bibinfo {author} {\bibfnamefont {T.~c.~v.}\ \bibnamefont {Stankevi\ifmmode~\check{c}\else \v{c}\fi{}}}, \bibinfo {author} {\bibfnamefont {L.}~\bibnamefont {Stek}}, \bibinfo {author} {\bibfnamefont {H.}~\bibnamefont {Suominen}}, \bibinfo {author} {\bibfnamefont {J.}~\bibnamefont {Suter}}, \bibinfo {author} {\bibfnamefont {V.}~\bibnamefont {Svidenko}}, \bibinfo {author} {\bibfnamefont {S.}~\bibnamefont {Teicher}}, \bibinfo {author} {\bibfnamefont {M.}~\bibnamefont {Temuerhan}}, \bibinfo {author} {\bibfnamefont {N.}~\bibnamefont {Thiyagarajah}}, \bibinfo {author} {\bibfnamefont {R.}~\bibnamefont {Tholapi}}, \bibinfo {author} {\bibfnamefont {M.}~\bibnamefont {Thomas}}, \bibinfo {author} {\bibfnamefont {E.}~\bibnamefont {Toomey}}, \bibinfo
  {author} {\bibfnamefont {S.}~\bibnamefont {Upadhyay}}, \bibinfo {author} {\bibfnamefont {I.}~\bibnamefont {Urban}}, \bibinfo {author} {\bibfnamefont {S.}~\bibnamefont {Vaitiek\ifmmode~\dot{e}\else \.{e}\fi{}nas}}, \bibinfo {author} {\bibfnamefont {K.}~\bibnamefont {Van~Hoogdalem}}, \bibinfo {author} {\bibfnamefont {D.}~\bibnamefont {Van~Woerkom}}, \bibinfo {author} {\bibfnamefont {D.~V.}\ \bibnamefont {Viazmitinov}}, \bibinfo {author} {\bibfnamefont {D.}~\bibnamefont {Vogel}}, \bibinfo {author} {\bibfnamefont {S.}~\bibnamefont {Waddy}}, \bibinfo {author} {\bibfnamefont {J.}~\bibnamefont {Watson}}, \bibinfo {author} {\bibfnamefont {J.}~\bibnamefont {Weston}}, \bibinfo {author} {\bibfnamefont {G.~W.}\ \bibnamefont {Winkler}}, \bibinfo {author} {\bibfnamefont {C.~K.}\ \bibnamefont {Yang}}, \bibinfo {author} {\bibfnamefont {S.}~\bibnamefont {Yau}}, \bibinfo {author} {\bibfnamefont {D.}~\bibnamefont {Yi}}, \bibinfo {author} {\bibfnamefont {E.}~\bibnamefont {Yucelen}}, \bibinfo {author} {\bibfnamefont
  {A.}~\bibnamefont {Webster}}, \bibinfo {author} {\bibfnamefont {R.}~\bibnamefont {Zeisel}},\ and\ \bibinfo {author} {\bibfnamefont {R.}~\bibnamefont {Zhao}} (\bibinfo {collaboration} {Microsoft Quantum}),\ }\bibfield  {title} {\bibinfo {title} {Inas-al hybrid devices passing the topological gap protocol},\ }\href {https://doi.org/10.1103/PhysRevB.107.245423} {\bibfield  {journal} {\bibinfo  {journal} {Phys. Rev. B}\ }\textbf {\bibinfo {volume} {107}},\ \bibinfo {pages} {245423} (\bibinfo {year} {2023})}\BibitemShut {NoStop}%
\bibitem [{\citenamefont {Kells}\ \emph {et~al.}(2012)\citenamefont {Kells}, \citenamefont {Meidan},\ and\ \citenamefont {Brouwer}}]{kells2012near}%
  \BibitemOpen
  \bibfield  {author} {\bibinfo {author} {\bibfnamefont {G.}~\bibnamefont {Kells}}, \bibinfo {author} {\bibfnamefont {D.}~\bibnamefont {Meidan}},\ and\ \bibinfo {author} {\bibfnamefont {P.}~\bibnamefont {Brouwer}},\ }\bibfield  {title} {\bibinfo {title} {Near-zero-energy end states in topologically trivial spin-orbit coupled superconducting nanowires with a smooth confinement},\ }\href@noop {} {\bibfield  {journal} {\bibinfo  {journal} {Physical Review B}\ }\textbf {\bibinfo {volume} {86}},\ \bibinfo {pages} {100503} (\bibinfo {year} {2012})}\BibitemShut {NoStop}%
\bibitem [{\citenamefont {Bagrets}\ and\ \citenamefont {Altland}(2012)}]{bagrets2012class}%
  \BibitemOpen
  \bibfield  {author} {\bibinfo {author} {\bibfnamefont {D.}~\bibnamefont {Bagrets}}\ and\ \bibinfo {author} {\bibfnamefont {A.}~\bibnamefont {Altland}},\ }\bibfield  {title} {\bibinfo {title} {Class d spectral peak in majorana quantum wires},\ }\href@noop {} {\bibfield  {journal} {\bibinfo  {journal} {Physical review letters}\ }\textbf {\bibinfo {volume} {109}},\ \bibinfo {pages} {227005} (\bibinfo {year} {2012})}\BibitemShut {NoStop}%
\bibitem [{\citenamefont {Pikulin}\ \emph {et~al.}(2012)\citenamefont {Pikulin}, \citenamefont {Dahlhaus}, \citenamefont {Wimmer}, \citenamefont {Schomerus},\ and\ \citenamefont {Beenakker}}]{pikulin2012zero}%
  \BibitemOpen
  \bibfield  {author} {\bibinfo {author} {\bibfnamefont {D.~I.}\ \bibnamefont {Pikulin}}, \bibinfo {author} {\bibfnamefont {J.}~\bibnamefont {Dahlhaus}}, \bibinfo {author} {\bibfnamefont {M.}~\bibnamefont {Wimmer}}, \bibinfo {author} {\bibfnamefont {H.}~\bibnamefont {Schomerus}},\ and\ \bibinfo {author} {\bibfnamefont {C.}~\bibnamefont {Beenakker}},\ }\bibfield  {title} {\bibinfo {title} {A zero-voltage conductance peak from weak antilocalization in a majorana nanowire},\ }\href@noop {} {\bibfield  {journal} {\bibinfo  {journal} {New Journal of Physics}\ }\textbf {\bibinfo {volume} {14}},\ \bibinfo {pages} {125011} (\bibinfo {year} {2012})}\BibitemShut {NoStop}%
\bibitem [{\citenamefont {Prada}\ \emph {et~al.}(2012)\citenamefont {Prada}, \citenamefont {San-Jose},\ and\ \citenamefont {Aguado}}]{prada2012transport}%
  \BibitemOpen
  \bibfield  {author} {\bibinfo {author} {\bibfnamefont {E.}~\bibnamefont {Prada}}, \bibinfo {author} {\bibfnamefont {P.}~\bibnamefont {San-Jose}},\ and\ \bibinfo {author} {\bibfnamefont {R.}~\bibnamefont {Aguado}},\ }\bibfield  {title} {\bibinfo {title} {Transport spectroscopy of n s nanowire junctions with majorana fermions},\ }\href@noop {} {\bibfield  {journal} {\bibinfo  {journal} {Physical Review B}\ }\textbf {\bibinfo {volume} {86}},\ \bibinfo {pages} {180503} (\bibinfo {year} {2012})}\BibitemShut {NoStop}%
\bibitem [{\citenamefont {Liu}\ \emph {et~al.}(2017)\citenamefont {Liu}, \citenamefont {Sau}, \citenamefont {Stanescu},\ and\ \citenamefont {Das~Sarma}}]{Liu_2017}%
  \BibitemOpen
  \bibfield  {author} {\bibinfo {author} {\bibfnamefont {C.-X.}\ \bibnamefont {Liu}}, \bibinfo {author} {\bibfnamefont {J.~D.}\ \bibnamefont {Sau}}, \bibinfo {author} {\bibfnamefont {T.~D.}\ \bibnamefont {Stanescu}},\ and\ \bibinfo {author} {\bibfnamefont {S.}~\bibnamefont {Das~Sarma}},\ }\bibfield  {title} {\bibinfo {title} {Andreev bound states versus majorana bound states in quantum dot-nanowire-superconductor hybrid structures: Trivial versus topological zero-bias conductance peaks},\ }\href {https://doi.org/10.1103/PhysRevB.96.075161} {\bibfield  {journal} {\bibinfo  {journal} {Phys. Rev. B}\ }\textbf {\bibinfo {volume} {96}},\ \bibinfo {pages} {075161} (\bibinfo {year} {2017})}\BibitemShut {NoStop}%
\bibitem [{\citenamefont {Pan}\ and\ \citenamefont {Sarma}(2020)}]{pan2020physical}%
  \BibitemOpen
  \bibfield  {author} {\bibinfo {author} {\bibfnamefont {H.}~\bibnamefont {Pan}}\ and\ \bibinfo {author} {\bibfnamefont {S.~D.}\ \bibnamefont {Sarma}},\ }\bibfield  {title} {\bibinfo {title} {Physical mechanisms for zero-bias conductance peaks in majorana nanowires},\ }\href@noop {} {\bibfield  {journal} {\bibinfo  {journal} {Physical Review Research}\ }\textbf {\bibinfo {volume} {2}},\ \bibinfo {pages} {013377} (\bibinfo {year} {2020})}\BibitemShut {NoStop}%
\bibitem [{\citenamefont {Moore}\ \emph {et~al.}(2018{\natexlab{a}})\citenamefont {Moore}, \citenamefont {Stanescu},\ and\ \citenamefont {Tewari}}]{moore2018two}%
  \BibitemOpen
  \bibfield  {author} {\bibinfo {author} {\bibfnamefont {C.}~\bibnamefont {Moore}}, \bibinfo {author} {\bibfnamefont {T.~D.}\ \bibnamefont {Stanescu}},\ and\ \bibinfo {author} {\bibfnamefont {S.}~\bibnamefont {Tewari}},\ }\bibfield  {title} {\bibinfo {title} {Two-terminal charge tunneling: Disentangling majorana zero modes from partially separated andreev bound states in semiconductor-superconductor heterostructures},\ }\href@noop {} {\bibfield  {journal} {\bibinfo  {journal} {Physical Review B}\ }\textbf {\bibinfo {volume} {97}},\ \bibinfo {pages} {165302} (\bibinfo {year} {2018}{\natexlab{a}})}\BibitemShut {NoStop}%
\bibitem [{\citenamefont {Moore}\ \emph {et~al.}(2018{\natexlab{b}})\citenamefont {Moore}, \citenamefont {Zeng}, \citenamefont {Stanescu},\ and\ \citenamefont {Tewari}}]{Moore2018}%
  \BibitemOpen
  \bibfield  {author} {\bibinfo {author} {\bibfnamefont {C.}~\bibnamefont {Moore}}, \bibinfo {author} {\bibfnamefont {C.}~\bibnamefont {Zeng}}, \bibinfo {author} {\bibfnamefont {T.~D.}\ \bibnamefont {Stanescu}},\ and\ \bibinfo {author} {\bibfnamefont {S.}~\bibnamefont {Tewari}},\ }\bibfield  {title} {\bibinfo {title} {Quantized zero-bias conductance plateau in semiconductor-superconductor heterostructures without topological majorana zero modes},\ }\href {https://doi.org/10.1103/PhysRevB.98.155314} {\bibfield  {journal} {\bibinfo  {journal} {Phys. Rev. B}\ }\textbf {\bibinfo {volume} {98}},\ \bibinfo {pages} {155314} (\bibinfo {year} {2018}{\natexlab{b}})}\BibitemShut {NoStop}%
\bibitem [{\citenamefont {Vuik}\ \emph {et~al.}(2018)\citenamefont {Vuik}, \citenamefont {Nijholt}, \citenamefont {Akhmerov},\ and\ \citenamefont {Wimmer}}]{vuik2018reproducing}%
  \BibitemOpen
  \bibfield  {author} {\bibinfo {author} {\bibfnamefont {A.}~\bibnamefont {Vuik}}, \bibinfo {author} {\bibfnamefont {B.}~\bibnamefont {Nijholt}}, \bibinfo {author} {\bibfnamefont {A.~R.}\ \bibnamefont {Akhmerov}},\ and\ \bibinfo {author} {\bibfnamefont {M.}~\bibnamefont {Wimmer}},\ }\bibfield  {title} {\bibinfo {title} {Reproducing topological properties with quasi-majorana states},\ }\href@noop {} {\bibfield  {journal} {\bibinfo  {journal} {arXiv preprint arXiv:1806.02801}\ } (\bibinfo {year} {2018})}\BibitemShut {NoStop}%
\bibitem [{\citenamefont {Stanescu}\ and\ \citenamefont {Tewari}(2019)}]{stanescu2019robust}%
  \BibitemOpen
  \bibfield  {author} {\bibinfo {author} {\bibfnamefont {T.~D.}\ \bibnamefont {Stanescu}}\ and\ \bibinfo {author} {\bibfnamefont {S.}~\bibnamefont {Tewari}},\ }\bibfield  {title} {\bibinfo {title} {Robust low-energy andreev bound states in semiconductor-superconductor structures: Importance of partial separation of component majorana bound states},\ }\href@noop {} {\bibfield  {journal} {\bibinfo  {journal} {Physical Review B}\ }\textbf {\bibinfo {volume} {100}},\ \bibinfo {pages} {155429} (\bibinfo {year} {2019})}\BibitemShut {NoStop}%
\bibitem [{\citenamefont {Reeg}\ \emph {et~al.}(2018)\citenamefont {Reeg}, \citenamefont {Dmytruk}, \citenamefont {Chevallier}, \citenamefont {Loss},\ and\ \citenamefont {Klinovaja}}]{added_Loss_2018prb_abs}%
  \BibitemOpen
  \bibfield  {author} {\bibinfo {author} {\bibfnamefont {C.}~\bibnamefont {Reeg}}, \bibinfo {author} {\bibfnamefont {O.}~\bibnamefont {Dmytruk}}, \bibinfo {author} {\bibfnamefont {D.}~\bibnamefont {Chevallier}}, \bibinfo {author} {\bibfnamefont {D.}~\bibnamefont {Loss}},\ and\ \bibinfo {author} {\bibfnamefont {J.}~\bibnamefont {Klinovaja}},\ }\bibfield  {title} {\bibinfo {title} {Zero-energy andreev bound states from quantum dots in proximitized rashba nanowires},\ }\href {https://doi.org/10.1103/PhysRevB.98.245407} {\bibfield  {journal} {\bibinfo  {journal} {Phys. Rev. B}\ }\textbf {\bibinfo {volume} {98}},\ \bibinfo {pages} {245407} (\bibinfo {year} {2018})}\BibitemShut {NoStop}%
\bibitem [{\citenamefont {San-Jose}\ \emph {et~al.}(2016)\citenamefont {San-Jose}, \citenamefont {Cayao}, \citenamefont {Prada},\ and\ \citenamefont {Aguado}}]{san2016majorana}%
  \BibitemOpen
  \bibfield  {author} {\bibinfo {author} {\bibfnamefont {P.}~\bibnamefont {San-Jose}}, \bibinfo {author} {\bibfnamefont {J.}~\bibnamefont {Cayao}}, \bibinfo {author} {\bibfnamefont {E.}~\bibnamefont {Prada}},\ and\ \bibinfo {author} {\bibfnamefont {R.}~\bibnamefont {Aguado}},\ }\bibfield  {title} {\bibinfo {title} {Majorana bound states from exceptional points in non-topological superconductors},\ }\href@noop {} {\bibfield  {journal} {\bibinfo  {journal} {Scientific reports}\ }\textbf {\bibinfo {volume} {6}},\ \bibinfo {pages} {21427} (\bibinfo {year} {2016})}\BibitemShut {NoStop}%
\bibitem [{\citenamefont {Avila}\ \emph {et~al.}(2019)\citenamefont {Avila}, \citenamefont {Peñaranda}, \citenamefont {Prada}, \citenamefont {San-Jose},\ and\ \citenamefont {Aguado}}]{ramon2019nonhermitian}%
  \BibitemOpen
  \bibfield  {author} {\bibinfo {author} {\bibfnamefont {J.}~\bibnamefont {Avila}}, \bibinfo {author} {\bibfnamefont {F.}~\bibnamefont {Peñaranda}}, \bibinfo {author} {\bibfnamefont {E.}~\bibnamefont {Prada}}, \bibinfo {author} {\bibfnamefont {P.}~\bibnamefont {San-Jose}},\ and\ \bibinfo {author} {\bibfnamefont {R.}~\bibnamefont {Aguado}},\ }\bibfield  {title} {\bibinfo {title} {Non-hermitian topology as a unifying framework for the andreev versus majorana states controversy},\ }\bibfield  {journal} {\bibinfo  {journal} {Communications Physics}\ }\textbf {\bibinfo {volume} {2}},\ \href {https://doi.org/10.1038/s42005-019-0231-8} {10.1038/s42005-019-0231-8} (\bibinfo {year} {2019})\BibitemShut {NoStop}%
\bibitem [{\citenamefont {Awoga}\ \emph {et~al.}(2019)\citenamefont {Awoga}, \citenamefont {Cayao},\ and\ \citenamefont {Black-Schaffer}}]{Jorge2019supercurrent}%
  \BibitemOpen
  \bibfield  {author} {\bibinfo {author} {\bibfnamefont {O.~A.}\ \bibnamefont {Awoga}}, \bibinfo {author} {\bibfnamefont {J.}~\bibnamefont {Cayao}},\ and\ \bibinfo {author} {\bibfnamefont {A.~M.}\ \bibnamefont {Black-Schaffer}},\ }\bibfield  {title} {\bibinfo {title} {Supercurrent detection of topologically trivial zero-energy states in nanowire junctions},\ }\href {https://doi.org/10.1103/PhysRevLett.123.117001} {\bibfield  {journal} {\bibinfo  {journal} {Phys. Rev. Lett.}\ }\textbf {\bibinfo {volume} {123}},\ \bibinfo {pages} {117001} (\bibinfo {year} {2019})}\BibitemShut {NoStop}%
\bibitem [{\citenamefont {Prada}\ \emph {et~al.}(2020)\citenamefont {Prada}, \citenamefont {San-Jose}, \citenamefont {de~Moor}, \citenamefont {Geresdi}, \citenamefont {Lee}, \citenamefont {Klinovaja}, \citenamefont {Loss}, \citenamefont {Nygård}, \citenamefont {Aguado},\ and\ \citenamefont {Kouwenhoven}}]{ramon2020from}%
  \BibitemOpen
  \bibfield  {author} {\bibinfo {author} {\bibfnamefont {E.}~\bibnamefont {Prada}}, \bibinfo {author} {\bibfnamefont {P.}~\bibnamefont {San-Jose}}, \bibinfo {author} {\bibfnamefont {M.~W.~A.}\ \bibnamefont {de~Moor}}, \bibinfo {author} {\bibfnamefont {A.}~\bibnamefont {Geresdi}}, \bibinfo {author} {\bibfnamefont {E.~J.~H.}\ \bibnamefont {Lee}}, \bibinfo {author} {\bibfnamefont {J.}~\bibnamefont {Klinovaja}}, \bibinfo {author} {\bibfnamefont {D.}~\bibnamefont {Loss}}, \bibinfo {author} {\bibfnamefont {J.}~\bibnamefont {Nygård}}, \bibinfo {author} {\bibfnamefont {R.}~\bibnamefont {Aguado}},\ and\ \bibinfo {author} {\bibfnamefont {L.~P.}\ \bibnamefont {Kouwenhoven}},\ }\bibfield  {title} {\bibinfo {title} {From andreev to majorana bound states in hybrid superconductor–semiconductor nanowires},\ }\bibfield  {journal} {\bibinfo  {journal} {Nature Reviews Physics}\ }\href {https://doi.org/10.1038/s42254-020-0228-y} {10.1038/s42254-020-0228-y} (\bibinfo {year} {2020})\BibitemShut {NoStop}%
\bibitem [{\citenamefont {Cayao}\ and\ \citenamefont {Black-Schaffer}(2021)}]{Jorge2021distinguishing}%
  \BibitemOpen
  \bibfield  {author} {\bibinfo {author} {\bibfnamefont {J.}~\bibnamefont {Cayao}}\ and\ \bibinfo {author} {\bibfnamefont {A.~M.}\ \bibnamefont {Black-Schaffer}},\ }\bibfield  {title} {\bibinfo {title} {Distinguishing trivial and topological zero-energy states in long nanowire junctions},\ }\bibfield  {journal} {\bibinfo  {journal} {Physical Review B}\ }\textbf {\bibinfo {volume} {104}},\ \href {https://doi.org/10.1103/physrevb.104.l020501} {10.1103/physrevb.104.l020501} (\bibinfo {year} {2021})\BibitemShut {NoStop}%
\bibitem [{\citenamefont {Turner}\ \emph {et~al.}(2011)\citenamefont {Turner}, \citenamefont {Pollmann},\ and\ \citenamefont {Berg}}]{Turner_2011}%
  \BibitemOpen
  \bibfield  {author} {\bibinfo {author} {\bibfnamefont {A.~M.}\ \bibnamefont {Turner}}, \bibinfo {author} {\bibfnamefont {F.}~\bibnamefont {Pollmann}},\ and\ \bibinfo {author} {\bibfnamefont {E.}~\bibnamefont {Berg}},\ }\bibfield  {title} {\bibinfo {title} {Topological phases of one-dimensional fermions: An entanglement point of view},\ }\href {https://doi.org/10.1103/PhysRevB.83.075102} {\bibfield  {journal} {\bibinfo  {journal} {Phys. Rev. B}\ }\textbf {\bibinfo {volume} {83}},\ \bibinfo {pages} {075102} (\bibinfo {year} {2011})}\BibitemShut {NoStop}%
\bibitem [{\citenamefont {Bonderson}\ \emph {et~al.}(2008)\citenamefont {Bonderson}, \citenamefont {Freedman},\ and\ \citenamefont {Nayak}}]{Bonderson_2008}%
  \BibitemOpen
  \bibfield  {author} {\bibinfo {author} {\bibfnamefont {P.}~\bibnamefont {Bonderson}}, \bibinfo {author} {\bibfnamefont {M.}~\bibnamefont {Freedman}},\ and\ \bibinfo {author} {\bibfnamefont {C.}~\bibnamefont {Nayak}},\ }\bibfield  {title} {\bibinfo {title} {Measurement-only topological quantum computation},\ }\href {https://doi.org/10.1103/PhysRevLett.101.010501} {\bibfield  {journal} {\bibinfo  {journal} {Phys. Rev. Lett.}\ }\textbf {\bibinfo {volume} {101}},\ \bibinfo {pages} {010501} (\bibinfo {year} {2008})}\BibitemShut {NoStop}%
\bibitem [{\citenamefont {Sau}\ \emph {et~al.}(2015)\citenamefont {Sau}, \citenamefont {Swingle},\ and\ \citenamefont {Tewari}}]{Sau_Tewari_2015}%
  \BibitemOpen
  \bibfield  {author} {\bibinfo {author} {\bibfnamefont {J.~D.}\ \bibnamefont {Sau}}, \bibinfo {author} {\bibfnamefont {B.}~\bibnamefont {Swingle}},\ and\ \bibinfo {author} {\bibfnamefont {S.}~\bibnamefont {Tewari}},\ }\bibfield  {title} {\bibinfo {title} {Proposal to probe quantum nonlocality of majorana fermions in tunneling experiments},\ }\href {https://doi.org/10.1103/PhysRevB.92.020511} {\bibfield  {journal} {\bibinfo  {journal} {Phys. Rev. B}\ }\textbf {\bibinfo {volume} {92}},\ \bibinfo {pages} {020511} (\bibinfo {year} {2015})}\BibitemShut {NoStop}%
\bibitem [{\citenamefont {Akhmerov}\ \emph {et~al.}(2009)\citenamefont {Akhmerov}, \citenamefont {Nilsson},\ and\ \citenamefont {Beenakker}}]{Akhmerov_2009}%
  \BibitemOpen
  \bibfield  {author} {\bibinfo {author} {\bibfnamefont {A.~R.}\ \bibnamefont {Akhmerov}}, \bibinfo {author} {\bibfnamefont {J.}~\bibnamefont {Nilsson}},\ and\ \bibinfo {author} {\bibfnamefont {C.~W.~J.}\ \bibnamefont {Beenakker}},\ }\bibfield  {title} {\bibinfo {title} {Electrically detected interferometry of majorana fermions in a topological insulator},\ }\href {https://doi.org/10.1103/PhysRevLett.102.216404} {\bibfield  {journal} {\bibinfo  {journal} {Phys. Rev. Lett.}\ }\textbf {\bibinfo {volume} {102}},\ \bibinfo {pages} {216404} (\bibinfo {year} {2009})}\BibitemShut {NoStop}%
\bibitem [{\citenamefont {Sau}\ \emph {et~al.}(2011)\citenamefont {Sau}, \citenamefont {Tewari},\ and\ \citenamefont {Das~Sarma}}]{Sau_Tewari_2011}%
  \BibitemOpen
  \bibfield  {author} {\bibinfo {author} {\bibfnamefont {J.~D.}\ \bibnamefont {Sau}}, \bibinfo {author} {\bibfnamefont {S.}~\bibnamefont {Tewari}},\ and\ \bibinfo {author} {\bibfnamefont {S.}~\bibnamefont {Das~Sarma}},\ }\bibfield  {title} {\bibinfo {title} {Probing non-abelian statistics with majorana fermion interferometry in spin-orbit-coupled semiconductors},\ }\href {https://doi.org/10.1103/PhysRevB.84.085109} {\bibfield  {journal} {\bibinfo  {journal} {Phys. Rev. B}\ }\textbf {\bibinfo {volume} {84}},\ \bibinfo {pages} {085109} (\bibinfo {year} {2011})}\BibitemShut {NoStop}%
\bibitem [{\citenamefont {{Microsoft Azure Quantum}}\ \emph {et~al.}(2025)\citenamefont {{Microsoft Azure Quantum}}, \citenamefont {Aghaee}, \citenamefont {Ramirez}, \citenamefont {Alam}, \citenamefont {Ali}, \citenamefont {Andrzejczuk}, \citenamefont {Antipov}, \citenamefont {Astafev}, \citenamefont {Barzegar}, \citenamefont {Bauer}, \citenamefont {Becker}, \citenamefont {Bhaskar}, \citenamefont {Bocharov}, \citenamefont {Boddapati}, \citenamefont {Bohn}, \citenamefont {Bommer}, \citenamefont {Bourdet}, \citenamefont {Bousquet}, \citenamefont {Boutin}, \citenamefont {Casparis}, \citenamefont {Chapman}, \citenamefont {...},\ and\ \citenamefont {Zilke}}]{Microsoft2025}%
  \BibitemOpen
  \bibfield  {author} {\bibinfo {author} {\bibnamefont {{Microsoft Azure Quantum}}}, \bibinfo {author} {\bibfnamefont {M.}~\bibnamefont {Aghaee}}, \bibinfo {author} {\bibfnamefont {A.~A.}\ \bibnamefont {Ramirez}}, \bibinfo {author} {\bibfnamefont {Z.}~\bibnamefont {Alam}}, \bibinfo {author} {\bibfnamefont {R.}~\bibnamefont {Ali}}, \bibinfo {author} {\bibfnamefont {M.}~\bibnamefont {Andrzejczuk}}, \bibinfo {author} {\bibfnamefont {A.}~\bibnamefont {Antipov}}, \bibinfo {author} {\bibfnamefont {M.}~\bibnamefont {Astafev}}, \bibinfo {author} {\bibfnamefont {A.}~\bibnamefont {Barzegar}}, \bibinfo {author} {\bibfnamefont {B.}~\bibnamefont {Bauer}}, \bibinfo {author} {\bibfnamefont {J.}~\bibnamefont {Becker}}, \bibinfo {author} {\bibfnamefont {U.~K.}\ \bibnamefont {Bhaskar}}, \bibinfo {author} {\bibfnamefont {A.}~\bibnamefont {Bocharov}}, \bibinfo {author} {\bibfnamefont {S.}~\bibnamefont {Boddapati}}, \bibinfo {author} {\bibfnamefont {D.}~\bibnamefont {Bohn}}, \bibinfo {author} {\bibfnamefont {J.}~\bibnamefont
  {Bommer}}, \bibinfo {author} {\bibfnamefont {L.}~\bibnamefont {Bourdet}}, \bibinfo {author} {\bibfnamefont {A.}~\bibnamefont {Bousquet}}, \bibinfo {author} {\bibfnamefont {S.}~\bibnamefont {Boutin}}, \bibinfo {author} {\bibfnamefont {L.}~\bibnamefont {Casparis}}, \bibinfo {author} {\bibfnamefont {B.~J.}\ \bibnamefont {Chapman}}, \bibinfo {author} {\bibnamefont {...}},\ and\ \bibinfo {author} {\bibfnamefont {J.}~\bibnamefont {Zilke}},\ }\bibfield  {title} {\bibinfo {title} {{Interferometric single-shot parity measurement in InAs–Al hybrid devices}},\ }\href@noop {} {\bibfield  {journal} {\bibinfo  {journal} {Nature}\ }\textbf {\bibinfo {volume} {638}},\ \bibinfo {pages} {651} (\bibinfo {year} {2025})}\BibitemShut {NoStop}%
\bibitem [{\citenamefont {Sau}\ and\ \citenamefont {Sarma}(2024)}]{Sau2024}%
  \BibitemOpen
  \bibfield  {author} {\bibinfo {author} {\bibfnamefont {J.~D.}\ \bibnamefont {Sau}}\ and\ \bibinfo {author} {\bibfnamefont {S.~D.}\ \bibnamefont {Sarma}},\ }\href@noop {} {\bibinfo {title} {{Capacitance-based Fermion parity read-out and predicted Rabi oscillations in a Majorana nanowire}}} (\bibinfo {year} {2024}),\ \bibinfo {note} {arXiv:2406.18080}\BibitemShut {NoStop}%
\bibitem [{\citenamefont {Eissele}\ \emph {et~al.}(2025)\citenamefont {Eissele}, \citenamefont {Roy}, \citenamefont {Tewari},\ and\ \citenamefont {Stanescu}}]{Eissele2025}%
  \BibitemOpen
  \bibfield  {author} {\bibinfo {author} {\bibfnamefont {R.}~\bibnamefont {Eissele}}, \bibinfo {author} {\bibfnamefont {B.}~\bibnamefont {Roy}}, \bibinfo {author} {\bibfnamefont {S.}~\bibnamefont {Tewari}},\ and\ \bibinfo {author} {\bibfnamefont {T.~D.}\ \bibnamefont {Stanescu}},\ }\href@noop {} {\bibinfo {title} {{Topological invariant for finite systems in the presence of disorder}}} (\bibinfo {year} {2025}),\ \bibinfo {note} {to be published}\BibitemShut {NoStop}%
\bibitem [{\citenamefont {Roy}\ \emph {et~al.}(2024)\citenamefont {Roy}, \citenamefont {Jaiswal}, \citenamefont {Stanescu},\ and\ \citenamefont {Tewari}}]{Roy_2024}%
  \BibitemOpen
  \bibfield  {author} {\bibinfo {author} {\bibfnamefont {B.~B.}\ \bibnamefont {Roy}}, \bibinfo {author} {\bibfnamefont {R.}~\bibnamefont {Jaiswal}}, \bibinfo {author} {\bibfnamefont {T.~D.}\ \bibnamefont {Stanescu}},\ and\ \bibinfo {author} {\bibfnamefont {S.}~\bibnamefont {Tewari}},\ }\bibfield  {title} {\bibinfo {title} {Stabilizing topological superconductivity in disordered spin-orbit coupled semiconductor-superconductor heterostructures},\ }\href {https://doi.org/10.1103/PhysRevB.110.115436} {\bibfield  {journal} {\bibinfo  {journal} {Phys. Rev. B}\ }\textbf {\bibinfo {volume} {110}},\ \bibinfo {pages} {115436} (\bibinfo {year} {2024})}\BibitemShut {NoStop}%
\end{thebibliography}%

\end{document}